\documentclass[12pt]{article}
\usepackage{amsmath}
\usepackage{amssymb}
\numberwithin{equation}{section}

\usepackage{epsf}
\usepackage{color}
\usepackage{graphicx}
\usepackage{amsmath}
\usepackage{amssymb}
\usepackage{latexsym}
\newcommand{\LamGUT}{\Lambda_{\mbox{{\scriptsize GUT}}}}
\newcommand{\msusy}{m_{\mbox{{\scriptsize SUSY}}}}
\newcommand{\etaCP}{\eta_{\mbox{{\scriptsize CP}}}}
\newcommand{\meff}{m_{\mbox{{\scriptsize eff}}}}
\newcommand{\VTB}{V_{\mbox{{\scriptsize TB}}}}

\begin{document}

\setlength{\baselineskip}{18pt}
\begin{titlepage}
\begin{flushright}
KYUSHU-HET-128
\end{flushright}

\vspace*{1.2cm}
\begin{center}
{\Large\bf Cascade Textures and SUSY {\boldmath $SO(10)$} GUT}
\end{center}
\lineskip .75em
\vskip 1.5cm

\begin{center}
{\large
Adisorn Adulpravitchai$^{a,}$\footnote[1]{E-mail:
\tt adisorn.adulpravitchai@mpi-hd.mpg.de},
Kentaro Kojima$^{b,}$\footnote[2]{E-mail:
\tt kojima@rche.kyushu-u.ac.jp},
and
Ryo Takahashi$^{a,}$\footnote[3]{E-mail:
\tt ryo.takahashi@mpi-hd.mpg.de }
}\\

\vspace{1cm}

$^a${\it Max-Planck-Institut f$\ddot{u}$r Kernphysik, Postfach 10 39 80,
69029 Heidelberg, Germany}
$^b${\it Center for Research and Advancement in Higher Education,\\
 Kyushu University, Fukuoka 810-8560, Japan}\\

\vspace*{10mm}
{\bf Abstract}\\[5mm]
{\parbox{13cm}{\hspace{5mm}
We give texture analyses of cascade hierarchical mass matrices in 
supersymmetric $SO(10)$ grand unified theory. We embed cascade 
mass textures of the standard model fermion with right-handed 
neutrinos into the theory, which gives relations among the mass 
matrices of the fermions. The related phenomenologies, such as the 
lepton flavor violating processes and leptogenesis, are also 
investigated in addition to the PMNS mixing angles.}}
\end{center}
\end{titlepage}

\newpage
\section{Introduction} 

The neutrino oscillation experiments have suggested that there are two 
large mixing angles among three generations in the lepton sector while 
all mixing angles in the quark sector are small. It is known that the 
current experimental data of leptonic mixing 
angles~\cite{Schwetz:2008er} is well approximated by the tri-bimaximal 
mixing~\cite{TBM}, which is given by¡¡
\begin{eqnarray}
  \VTB=
  \left(
    \begin{array}{ccc}
      2/\sqrt{6}  & 1/\sqrt{3} & 0           \\
      -1/\sqrt{6} & 1/\sqrt{3} & -1/\sqrt{2} \\
      -1/\sqrt{6} & 1/\sqrt{3} & 1/\sqrt{2}
    \end{array}
  \right). \label{VTB}
\end{eqnarray}
Such suggestive form of the generation mixing gives us a strong 
motivation to study a flavor structure of the lepton sector. Actually, 
there are a number of proposals based on a flavor symmetry to unravel 
it and related phenomenologies have been elaborated~\cite{Altarelli:2005yp}. 

It has been pointed out that the neutrino Dirac mass matrix of the 
cascade form can lead to the tri-bimaximal mixing at the leading order 
in the framework of type I seesaw mechanism~\cite{Haba:2008dp}. The 
mass matrix of the cascade form is parametrized by 
\begin{eqnarray}
  M_{\mbox{{\scriptsize cas}}} \simeq
   \left(
    \begin{array}{ccc}
      \delta & \delta  & \delta \\
      \delta & \lambda & \lambda \\
      \delta & \lambda & 1
    \end{array}
   \right)v,~~~\mbox{ with }~~~|\delta|\ll|\lambda|\ll1,
  \label{cas}
\end{eqnarray}
and $v$ denotes an overall mass scale. We call this kind of hierarchy 
and the matrix with such hierarchy, ``cascade hierarchy'' and 
``cascade matrix'', respectively. On the other hand, the down quark 
mass matrix of a different hierarchical form, which is 
\begin{eqnarray}
  M_{\mbox{{\scriptsize hyb}}} \simeq
  \left(
    \begin{array}{ccc}
      \epsilon' & \delta' & \delta' \\
      \delta'   & \lambda' & \lambda' \\
      \delta'   & \lambda' & 1
    \end{array}
  \right)v',~~~\mbox{ with }~~~|\epsilon'|\ll|\delta'|\ll|\lambda'|\ll1,
  \label{hyb}
\end{eqnarray}
can give realistic values of CKM matrix elements. The $(1,1)$ element, 
$\epsilon'$, of this matrix is smaller than all other elements but 
this hierarchical structure is close to the cascade form except for 
$\epsilon'$. We call this type of hierarchy ``hybrid cascade (H.C.) 
hierarchy'', and the matrix with such a hierarchy ``hybrid cascade 
(H.C.) matrix''. The neutrino Dirac mass matrix of a cascade form 
gives nearly tri-bimaximal generation mixing and the down quark mass 
matrix of a H.C. form realizes the CKM structure. The fact gives us a 
strong motivation to comprehensively investigate the quark and 
lepton. Actually, a proposal to embed such cascade textures into a 
supersymmetric (SUSY) $SU(5)$ grand unified theory and texture 
analyses have been presented~\cite{su5}. For comparison with a $SU(5)$ 
case and its results, we investigate embedding cascade hierarchies 
into a SUSY $SO(10)$ GUT in this paper, which is also one of 
fascinating grand unified models. 

The paper is organized as follows: In section 2, we give a brief 
review of cascade hierarchies for the fermion masses and mixing 
angles. In section 3, we embed the cascade hierarchies into the SUSY 
$SO(10)$ GUT. The texture analyses for the quark/lepton sectors are 
also given in the section. In section 4, we show some numerical 
analyses of our model. In section 5, we give a comment on the proton 
decay. Section 6 is devoted to the summary. Appendix A gives a 
discussion about constraints on the structure of right-handed neutrino 
mass matrix. 

\section{Cascade hierarchies for fermion mass matrices}

In this section, we give a brief review of cascade hierarchies for 
mass matrices of the fermions. First we discuss the cascade textures 
for quark and lepton sectors independently. The study of cascades for 
the lepton sector has been discussed in~\cite{Haba:2008dp}. Then a 
possible expansion of the study to quark sector was presented 
in~\cite{su5}, which was considered in a SUSY $SU(5)$ GUT. The 
work~\cite{Haba:2008dp} has pointed out that the neutrino Dirac mass 
matrix of a cascade form can lead to the tri-bimaximal mixing at the 
leading order in the framework of type I seesaw mechanism. Since the 
tri-bimaximal structure can be almost induced from the neutrino 
sector, the mixing angles from the charged lepton sectors should be 
small~\cite{su5}. This means that the form of charged lepton mass 
matrix can be taken as a cascade or H.C. because mixing angles for 
those textures are small enough. How about the quark sector? The CKM 
structure is almost determined by a structure of down quark mass 
matrix because of large mass hierarchies among up-type quarks. And it 
is known that the down quark mass matrix of a H.C. form can give the 
realistic CKM structure (e.g. see~\cite{Haba:2008dp,su5}). The 
contributions from up-quark sector to the CKM mixing are automatically 
tiny. This means that the form of up quark mass matrix can be taken as 
a cascade or H.C.. Finally, we comment on the structure of 
right-handed Majorana mass matrix. The contribution from the 
right-handed Majorana mass matrix should be also small because a 
nearly tri-bimaximal mixing are almost induced from the neutrino Dirac 
mass matrix with the seesaw mechanism, which means it is possible to 
take the right-handed Majorana mass matrix as a cascade or H.C.. More 
detailed explanations about the above points including mass 
eigenvalues given from each mass matrix of cascade and H.C. is given 
in~\cite{su5}. Here, we summarize the above discussions about possible 
structures of mass matrices of the fermions as, 
\begin{eqnarray}
  M_u &:& \mbox{cascade or H.C. or small mixing matrix}, \\
  M_d &:& \mbox{H.C.}, \\
  M_{\nu D} &:& \mbox{cascade}, \\
  M_e &:& \mbox{cascade or H.C. or small mixing matrix}, \\
  M_R &:& \mbox{cascade or H.C. or small mixing matrix}, 
\end{eqnarray}
where $M_u$, $M_d$, $M_{\nu D}$, $M_e$, and $M_R$ are mass matrices of 
up- and down-type quarks, neutrino Dirac, charged lepton, and 
right-handed neutrino, respectively. 

Next, we comment on the cascade textures in a $SU(5)$ case. The 
$SU(5)$ GUT predicts a relation between mass matrices for the 
down-type quark and charged lepton, 
\begin{eqnarray}
  M_e\simeq M_d^T, \label{su5rel}
\end{eqnarray}
due to an unification of matter contents. As discussed above, since 
only mass matrix of a H.C. form are allowed for $M_d$ in the study of 
cascade texture, the mass matrix for charged lepton should also have 
the H.C. from. On the other hand, some hierarchical structure of the 
mass matrices for the up-type quark and right-handed neutrino are 
allowed as long as induced mixing angles from these matrices can be 
treated as collections for the CKM and PMNS structures, 
respectively. Therefore, we can parametrize the mass matrices of the 
cascade or H.C. form for the fermions as 
\begin{eqnarray}
  M_u &\simeq& 
  \left(
    \begin{array}{ccc}
      \epsilon_u & \delta_u  & \delta_u \\
      \delta_u & \lambda_u & \lambda_u \\
      \delta_u & \lambda_u & 1
    \end{array}
  \right)v_u,\hspace{3mm}\mbox{ with }
  \left\{
    \begin{array}{l}
      |\epsilon_u|=|\delta_u|\ll|\lambda_u|\ll1:\mbox{cascade}, \\
      |\epsilon_u|\ll|\delta_u|\ll|\lambda_u|\ll1:\mbox{H.C.},
    \end{array}
  \right.
  \\
  M_d &\simeq& 
  \left(
    \begin{array}{ccc}
      \epsilon_d & \delta_d  & \delta_d \\
      \delta_d   & \lambda_d & \lambda_d \\
      \delta_d   & \lambda_d & 1
    \end{array}
  \right)\xi_dv_d,\hspace{0.5mm}\mbox{ with }|\epsilon_d|\ll|\delta_d|\ll|\lambda_d|\ll1:\mbox{H.C.},
  \\
  M_{\nu D} &\simeq&
  \left(
    \begin{array}{ccc}
      \delta_\nu & \delta_\nu  & \delta_\nu \\
      \delta_\nu & \lambda_\nu & \lambda_\nu \\
      \delta_\nu & \lambda_\nu & 1
    \end{array}
  \right)\xi_\nu v_u,\mbox{ with } |\delta_\nu|\ll|\lambda_\nu|\ll1:\mbox{cascade}, 
  \\
  M_e &\simeq& 
  \left(
    \begin{array}{ccc}
      \epsilon_d & \delta_d   & \delta_d  \\
      \delta_d   & -3\lambda_d & \lambda_d \\
      \delta_d   & \lambda_d & 1
    \end{array}
  \right)\xi_dv_d,\hspace{1mm}\mbox{ with } |\epsilon_d|\ll|\delta_d|\ll|\lambda_d|\ll1:\mbox{H.C.},
\end{eqnarray}
without $\mathcal{O}(1)$ coefficients for all elements. Here $v_u$ and 
$v_d$ are vacuum expectation values (VEVs) of up- and down-type Higgs 
fields in a supersymmetric scenario, and the overall factor $\xi_d$ 
and $\xi_\nu$ could be small. We also notice that the Georgi-Jarlskog 
(GJ) factor~\cite{Georgi:1979df} is introduced to mass ratio between 
the down-type quarks and charged leptons for each generation, 
\begin{eqnarray}
  \left(\frac{m_\tau}{m_b},\frac{m_\mu}{m_s},\frac{m_e}{m_d}\right)
  \sim\left(1,3,\frac{1}{3}\right).\label{GJrel}
\end{eqnarray}

\section{Cascade hierarchies in {\boldmath$SO(10)$} GUT}

\subsection{A SUSY {\boldmath $SO(10)$} Model}

We consider about embedding the (hybrid) cascade hierarchical mass 
matrices into $SO(10)$ GUT in this paper. A simple $SO(10)$ GUT 
predicts relations between mass matrices the up-type quark and 
neutrino Dirac, 
\begin{eqnarray}
  M_u\simeq M_{\nu D},
\end{eqnarray}
in addition to the relation~\eqref{su5rel}. As discussed above, since 
only the mass matrix of the H.C. form is allowed for $M_d$, the mass 
matrix for the charged lepton should also have the H.C. form like in 
$SU(5)$ case. For the up-type quark sector, a simple $SO(10)$ case 
predicts a GUT relation of the mass matrices $M_u\simeq M_{\nu D}$, 
and the up-type quark mass matrix $M_u$ should be restricted to a 
cascade form because the cascade form of neutrino Dirac mass matrix is 
needed for generating the tri-bimaximal neutrino mixing at the leading 
order. For the structures of right-handed neutrino mass matrix, some 
arbitrary matrices are allowed as long as induced mixing angles can be 
treated as collections for the PMNS matrix. 

To demonstrate the idea, we consider a simple SUSY $SO(10)$ model, 
which the Standard Model (SM) fermions with the right-handed neutrino 
are included into the spinor $16$-dimensional representation, $\psi$. 
To give suitable fermion masses, we introduce the Higgs fields, 
i.e. two Higgs $10$-plets, $H_{1,2}$ and two Higgs $\bar{126}$-plets, 
$\bar{\Delta}_{1,2}$. There are several ways to break $SO(10)$ down to the SM.
Here, we consider a minimal framework where the breaking of $SO(10)$ is achieved by the Higgs $210$-plet~\cite{Aulakh:1982sw,Clark:1982ai}, $\Phi$, which breaks $SO(10)$ down to Pati-Salam group: $SU(4)_C \times SU(2)_L \times SU(2)_R$. We choose that the Pati-Salam group is broken further down to the SM via the VEV of the SM singlet component in $\bar{\Delta}_2$ and the VEV also gives Majorana masses for the right-handed neutrinos. Since this singlet VEV gives the non-vanishing contribution to D-term in the superpotential resulting in the unwanted source of SUSY breaking at high energy (close to the GUT scale), we introduce a $126$-plet, $\Delta$, whose SM singlet component obtains the VEV to cancel the D-term contribution~(for instance, see~\cite{Aulakh:2003kg,Bajc:2004xe}).
Here we include two $10$-plets, $H_{1,2}$ because the 
mass matrices of the up-type quark and down-type quark have to be 
different in order to predict the correct CKM mixing angles, as well 
motivated from the previous discussion. Moreover, we also need one 
$\bar{126}$-plet, $\bar{\Delta}_1$, in order to achieve the GJ relations~\eqref{GJrel}, 
that is, to give the factor of $-3$ in the (2,2), (2,3), and (3,2) 
elements of the charged lepton mass matrix with respect to that of the 
down quark mass matrix.\footnote{ We note that the factor of $-3$ can 
  be obtained by the coupling of the Higgs $120$ or $\bar{126}$, see 
  for instance~\cite{Chen:2003zv}.} Another $\bar{126}$-plet, $\bar{\Delta}_2$, is 
introduced to generate the different texture for the right-handed neutrino masses and also break the Pati-Salam group to the SM. In our setup, there are six pairs of Higgs doublets, $\phi_u=(H_{1,u},H_{2,u},\bar{\Delta}_{1,u},\bar{\Delta}_{2,u},\Delta_{u},\Phi_u)^T$ and $\phi_d=(H_{1,d},H_{2,d},\bar{\Delta}_{1,d},\bar{\Delta}_{2,d},\Delta_{d},\Phi_d)^T$ with the mass term $\phi_u M_{H} \phi_d^T$. Note the label $u,d$ refer to the $SU(2)_L$ doublet component with hypercharge $\pm 1/2$ within the GUT multiplet. The mass matrix $M_H$ can be diagonalized by $U_{\phi_u}^T M_H U_{\phi_d}$, which $U_{\phi_u},U_{\phi_d}$ are unitaty matrices acting on $\phi_u$ and $\phi_d$ respectively. In the diagonal basis, the Higgs fields are given by $(\phi_u')_\alpha=(U_{\phi_u}^*)_{\beta \alpha} (\phi_u)_\beta$ and $(\phi_d')_\alpha=(U_{\phi_d}^*)_{\beta \alpha} (\phi_d)_\beta$. For the sake of the study, we will not specify how $SO(10)$ is broken in detail, but by 
some doublet-triplet splitting mechanism~(for instance 
see~\cite{Dimopoulos:1981xm,Chacko:1998zn,Lee:1994je}) we will assume that $H_u=(\phi_u')_1$ and $H_d=(\phi_d')_1$ have mass at the electroweak scale while the others are so heavy and decoupled from the low energy theory. The Higgs fields, $H_{u,d}$ are the two Higgs doublets of the Minimal 
Supersymmetric Standard Model (MSSM).

The superpotential of the model is given by
\begin{equation}
  W_{Y} = \tilde{Y}^{10}_{1} \psi H_1 \psi + \tilde{Y}^{10}_{2} \psi H_2 \psi +
  \tilde{Y}^{\bar{126}}_{1} \psi \bar{\Delta}_1 \psi +\tilde{Y}^{\bar{126}}_2 \psi
  \bar{\Delta}_2 \psi\;,
\end{equation}
which can be written in terms of the SM components as follows~\cite{Fukuyama:2004ps}:
\begin{eqnarray}
 W_{Y} &\ni& Q (\tilde{Y}^{10}_{1} H_{1,u} +\tilde{Y}^{10}_{2} H_{2,u} + \tilde{Y}^{\bar{126}}_{1} \bar{\Delta}_{1,u} + \tilde{Y}^{\bar{126}}_{2} \bar{\Delta}_{2,u}) U^c \nonumber \\ 
 & &+L (\tilde{Y}^{10}_{1} H_{1,u} +\tilde{Y}^{10}_{2} H_{2,u} -3 \tilde{Y}^{\bar{126}}_{1} \bar{\Delta}_{1,u} -3 \tilde{Y}^{\bar{126}}_{2} \bar{\Delta}_{2,u}) N \nonumber \\ 
 & &+Q (\tilde{Y}^{10}_{1} H_{1,d} +\tilde{Y}^{10}_{2} H_{2,d} + \tilde{Y}^{\bar{126}}_{1} \bar{\Delta}_{1,d} + \tilde{Y}^{\bar{126}}_{2} \bar{\Delta}_{2,d}) D^c \nonumber \\ 
 & &+L (\tilde{Y}^{10}_{1} H_{1,d} +\tilde{Y}^{10}_{2} H_{2,d} -3 \tilde{Y}^{\bar{126}}_{1} \bar{\Delta}_{1,d} -3 \tilde{Y}^{\bar{126}}_{2} \bar{\Delta}_{2,d}) E^c \;,
\end{eqnarray}
where the doublet component in the GUT multiplet can be written in term of the MSSM Higgs doublets as $(\phi_u)_\alpha=(U_{\phi_u})_{\alpha 1} H_u$ and $(\phi_d)_\alpha=(U_{\phi_d})_{\alpha 1} H_d$. 

For the neutrino sector, we assume that the $SU(2)_L$ triplet component, $\bar{\Delta}_{2,T}$, and the SM singlet component, $\bar{\Delta}_{2,S}$, in $\bar{\Delta}_2$, give tiny Majorana masses for the left-handed neutrinos and the heavy Majorana masses for the right-handed neutrinos respectively. This results in the seesaw formula as follow:  
\begin{equation}
 M_{\nu} = M_{LL} - M_{\nu D}^T M_{R}^{-1} M_{\nu D} \;, \label{SO10: seesaw}
\end{equation}
where $M_{LL}=\tilde{Y}^{\bar{126}}_{2} \langle \bar{\Delta}_{2,T} \rangle=\tilde{Y}^{\bar{126}}_{2} v_L$, $M_{R}=\tilde{Y}^{\bar{126}}_{2} \langle \bar{\Delta}_{2,S} \rangle=\tilde{Y}^{\bar{126}}_{2} v_R$ and $M_{\nu D}$ is the Dirac mass term whose structure will be discussed below. 
Since the triplet VEV $\langle \bar{\Delta}_{2,T} \rangle=v_L$ depends on parameters in Higgs superpotential~(for instance, see~\cite{Aulakh:2003kg}), we assume that the VEV is tiny such that the second term in Eq~(\ref{SO10: seesaw}) dominates, resulting in the type I seesaw dominance. Note that the singlet VEV, $\langle \bar{\Delta}_{2,S} \rangle=v_R$, is of order $10^{16}$~GeV. 

After the electroweak symmetry is broken via the doublet VEVs, $\langle H_{u,d} \rangle=v_{u,d}$, the fermion masses are given by 
\begin{eqnarray}
 M_u &\simeq&  (U_{\phi_u})_{1 1} \tilde{Y}^{10}_1 v_u = Y^{10}_1 v_u\\
 M_{\nu_D} &\simeq& (U_{\phi_u})_{1 1} \tilde{Y}^{10}_1 v_u = Y^{10}_1 v_u \\
 M_d &\simeq& ((U_{\phi_d})_{2 1} \tilde{Y}^{10}_2 + (U_{\phi_d})_{3 1} \tilde{Y}^{126}_1) v_d = ( Y^{10}_2 +  Y^{126}_1) v_d \\
 M_e &\simeq& ((U_{\phi_d})_{2 1} \tilde{Y}^{10}_2 -3 (U_{\phi_d})_{3 1} \tilde{Y}^{126}_1) v_d =( Y^{10}_2 -3  Y^{126}_1) v_d\;, 
\end{eqnarray}
where we assume that the main contribution for the up-type quark (Dirac 
neutrino) masses comes from the coupling to $H_1$ while for the down-type quark
 (charged lepton) masses they arise from the $H_2$ and $\bar{\Delta}_1$ 
couplings. These can be achieved through the following assumptions: 
$(U_{\phi_u})_{11}\gg(U_{\phi_u})_{i1}$ and 
$(U_{\phi_d})_{21},(U_{\phi_d})_{31}\gg(U_{\phi_d})_{i1}$. The Yukawa couplings
 are defined as $Y^{10}_1=(U_{\phi_u})_{1 1} \tilde{Y}^{10}_1$, 
$Y^{10}_2=(U_{\phi_d})_{2 1} \tilde{Y}^{10}_2$ and 
$Y^{126}_1=(U_{\phi_d})_{3 1} \tilde{Y}^{126}_1$. We impose the hierarchical 
forms to the Yukawa couplings, 
\begin{eqnarray}
  Y^{10}_1 &\simeq&
  \left(
    \begin{array}{ccc}
      \delta_u & \delta_u  & \delta_u \\
      \delta_u & \lambda_u & \lambda_u \\
      \delta_u & \lambda_u & 1
    \end{array}
  \right),\hspace{3mm}\mbox{ with }
  |\delta_u|\ll|\lambda_u|\ll1  , \\
  Y^{10}_2 &\simeq&
  \left(
    \begin{array}{ccc}
      \epsilon_d & \delta_d  & \delta_d \\
      \delta_d   & \delta_d & \delta_d \\
      \delta_d   & \delta_d & 1
    \end{array}
  \right),\\
  Y^{\bar{126}}_1 &\simeq&
  \left(
    \begin{array}{ccc}
      0 & 0  & 0 \\
      0   & \lambda_d & \lambda_d \\
      0   & \lambda_d & \lambda_d
    \end{array}
  \right),\hspace{0.5mm}\mbox{ with }
  |\epsilon_d|\ll|\delta_d|\ll|\lambda_d|\ll1 .
\end{eqnarray} 
The structure of $\tilde{Y}^{\bar{126}}_2$ will be discussed in the following sections in term of the right-handed neutrino mass matrix, $M_R=\tilde{Y}^{\bar{126}}_2 v_R$. These lead to the fermion mass matrices at the GUT scale as 
\begin{eqnarray}
  M_u &\simeq&
  \left(
    \begin{array}{ccc}
      \delta_u & \delta_u  & \delta_u \\
      \delta_u & \lambda_u & \lambda_u \\
      \delta_u & \lambda_u & 1
    \end{array}
  \right)v_u,\label{Renorm1}\\
  M_{\nu D} &\simeq&
  \left(
    \begin{array}{ccc}
      \delta_u & \delta_u  & \delta_u \\
      \delta_u & \lambda_u & \lambda_u \\
      \delta_u & \lambda_u & 1
    \end{array}
  \right) v_u,\hspace{3mm}\mbox{ with }
  |\delta_u|\ll|\lambda_u|\ll1  :  \mbox{cascade},  \label{Renorm3}\\
  M_d &\simeq&
  \left(
    \begin{array}{ccc}
      \epsilon_d & \delta_d  & \delta_d \\
      \delta_d   & \lambda_d & \lambda_d \\
      \delta_d   & \lambda_d & 1
    \end{array}
  \right) \xi_dv_d,\label{Renorm2} \\
  M_e &\simeq&
  \left(
    \begin{array}{ccc}
      \epsilon_d & \delta_d  & \delta_d  \\
      \delta_d   & -3\lambda_d & -3\lambda_d \\
      \delta_d   & -3\lambda_d & 1
    \end{array}
  \right)\xi_dv_d,\hspace{0.5mm}\mbox{ with }
  |\epsilon_d|\ll|\delta_d|\ll|\lambda_d|\ll1:\mbox{H.C.},\label{Renorm4}
\end{eqnarray}
where $\mathcal{O}(1)$ coefficients for all elements have been dropped. 
\subsection{Cabibbo fitting of cascade mass matrices}
The cascade hierarchical parameters are determined by observed 
values. It is naturally expected that such hierarchies are originated 
from a symmetry and/or some dynamics in a high energy regime rather 
than solely determined by the magnitudes of Yukawa couplings. 
Although the origin of the hierarchies is not specified in the 
analysis, one can estimate and study the relative magnitudes of the 
hierarchies introducing a small parameter in the mass matrices. In 
the following, we choose the Cabibbo angle, 
$\sin\theta_c\simeq\lambda=0.227$, as a fitting parameter, and study 
significant implications of the cascade $SO(10)$ scenario. Then we have 
\begin{eqnarray}
  \lambda_u\simeq0.87\times\lambda^4,~~~\delta_u\simeq0.85\times\lambda^8,
\end{eqnarray}
for up-quark mass matrix of the cascade form and
\begin{eqnarray}
  \lambda_d\simeq0.35\times\lambda^2,~~~\delta_d\simeq0.35\times\lambda^3,
\end{eqnarray}
for down-quark one of the H.C. form at GUT scale, where we utilized values of
quark masses listed in~\cite{Ross:2007az}.

Notice that the $\xi_d$ is a parameter, which determines a ratio 
between (3,3) element of Yukawa matrices for up- and down-type quarks, 
and thus, it is correlated with the $\tan\beta$ as, 
\begin{eqnarray}
  \tan\beta=\frac{v_u}{v_d}\simeq
  \left\{
    \begin{array}{llll}
      \phantom{\lambda^2}m_t/m_b\sim\mathcal{O}(50) & \mbox{ for } &
      \xi_d\sim\lambda^0 & \mbox{ [large]}\\
      \phantom{^2}\lambda m_t/m_b\sim\mathcal{O}(10) & \mbox{ for } &
      \xi_d\sim\lambda^1 & \mbox{ [moderate]}\\
      \lambda^2m_t/m_b\sim\mathcal{O}(1) & \mbox{ for } &
      \xi_d\sim\lambda^2 & \mbox{ [small]}
    \end{array}
  \right..
\end{eqnarray}
As the results we can write cascading textures at GUT scale as 
\begin{eqnarray}
  M_u &\simeq&
  \left(
    \begin{array}{ccc}
      \lambda^8 & \lambda^8 & \lambda^8 \\
      \lambda^8       & \lambda^4 & \lambda^4 \\
      \lambda^8       & \lambda^4 & 1
    \end{array}
  \right)v_u, \label{Mu-cas} \\
  M_d &\simeq&
  \left\{
    \begin{array}{ll}
     \left(
       \begin{array}{ccc}
         \lambda^{k_d+3} & \lambda^3 & \lambda^3 \\
         \lambda^3       & \lambda^2 & \lambda^2 \\
         \lambda^3       & \lambda^2 & 1
      \end{array}
    \right)v_d & \mbox{[large $\tan\beta$]} \\
    \left(
      \begin{array}{ccc}
        \lambda^{k_d+4} & \lambda^4 & \lambda^4 \\
        \lambda^4       & \lambda^3 & \lambda^3 \\
        \lambda^4       & \lambda^3 & \lambda
      \end{array}
    \right)v_d & \mbox{[moderate $\tan\beta$]} \\
    \left(
      \begin{array}{ccc}
        \lambda^{k_d+5} & \lambda^5 & \lambda^5 \\
        \lambda^5       & \lambda^4 & \lambda^4 \\
        \lambda^5       & \lambda^4 & \lambda^2
      \end{array}
    \right)v_d & \mbox{[small $\tan\beta$]} \\
    \end{array}
  \right.,\label{Md-cas}
\end{eqnarray}
where $k_d\geq1$ is needed to obtain suitable mass eigenvalues after 
diagonalizing these matrices. It should be remembered that $M_e\sim 
M_d$ but the additional GJ factor $-3$ is multiplied to the (2,2), 
(2,3), and (3,2) elements of $M_e$ as discussed in the previous 
section. 

\subsection{Neutrino sector}

Next, we consider the structure of neutrino mass matrices. In the 
cascade model~\cite{Haba:2008dp,su5}, cascade parameters are 
constrained as 
\begin{eqnarray}
  \left|\frac{\delta_\nu}{\lambda_\nu}\right|^2
  \ll\frac{\Delta m_{21}^2}{|\Delta m_{31}^2|}
  \simeq3.19\times10^{-2}<\lambda^2,
  \label{constraint}
\end{eqnarray}
in order to preserve the tri-bimaximal mixing at the leading order with 
\begin{eqnarray}
  \Delta m_{21}^2 &=& (7.695\pm0.645)\times10^{-5}\mbox{ eV}^2, \label{sol}\\
  |\Delta m_{31}^2| &=& 2.40_{-0.11}^{+0.12}\times10^{-3}\mbox{ eV}^2.
\label{atm}
\end{eqnarray}
at the $3\sigma$ level~\cite{Schwetz:2008er}. Due to the $SO(10)$ 
GUT relation $M_{\nu D}\simeq M_u$, the neutrino mass matrix can be 
parametrized as 
\begin{eqnarray}
  M_{\nu D} &\simeq&
  \left(
    \begin{array}{ccc}
      \lambda^8 & \lambda^8 & \lambda^8 \\
      \lambda^8 & \lambda^4 & -\lambda^4 \\
      \lambda^8 & -\lambda^4 & 1
    \end{array}
  \right)v_u \;, 
  \label{Dirac}
\end{eqnarray}
where we note that an opposite sign between (2,2) and (2,3) elements 
is experimentally required to obtain the tri-bimaximal mixing as 
commented in~\cite{Haba:2008dp}.\footnote{ Since the Dirac mass matrix 
  and the up-type quark mass matrix are constrained to have the same
  structure, this opposite sign is also imposed on the up-type quark
  mass matrix~\eqref{Mu-cas}.}

\subsubsection{Diagonal {\boldmath $M_R$} case}

Let us discuss the case of a diagonal Majorana mass matrix of the 
right-handed neutrinos, 
$M_R=\mbox{Diag}[\lambda^{x_1},\lambda^{x_2},1]M$, where $x_1\geq 
x_2\geq0$. The cascade model requires the normal mass hierarchy of 
light neutrino mass spectrum in order to realize a nearly 
tri-bimaximal mixing~\cite{Haba:2008dp}. The mass eigenvalues can 
be estimated as 
\begin{eqnarray}
  m_1&\simeq&\frac{
    v_u^2}{6M}\equiv\bar{m}_1, \label{m1}\\
  m_2&\simeq&\left(3\lambda^{16-x_1}+\frac{1}{3}\right)
  \frac{
    v_u^2}{M}\equiv\bar{m}_2+2\bar{m_1}, \label{m2}\\
  m_3&\simeq&\left(2\lambda^{8-x_2}+\frac{1}{2}\right)
  \frac{
    v_u^2}{M}\equiv\bar{m}_3+3\bar{m_3},
  \label{m3}
\end{eqnarray}
with a leading order corrections of $\mathcal{O}(\bar{m}_1)$. In order 
to understand the hierarchical structure of the mass matrix and the 
constraints on the cascade parameters, we write down the effective 
neutrino mass matrix as 
\begin{eqnarray}
  M_\nu &\simeq& \frac{
    v_u^2}{M}
  \left(
    \begin{array}{ccc}
      4  & -2 & -2 \\
      -2 & 1  & 1  \\
      -2 & 1  & 1
    \end{array}
  \right)+\frac{\lambda^{
      16-x_1}v_u^2}{M}
  \left(
    \begin{array}{ccc}
      1 & 1 & 1 \\
      1 & 1 & 1 \\
      1 & 1 & 1
    \end{array}
  \right)
  +\frac{\lambda^{
      8-x_2}v_u^2}{M}
  \left(
    \begin{array}{ccc}
      0 & 0  & 0  \\
      0 & 1  & -1 \\
      0 & -1 & 1
    \end{array}
  \right) \nonumber 
  \\
  &&
  +\frac{
    v_u^2}{M}
  \left(
    \begin{array}{ccc}
      -4+\lambda^{16}   & 2-\lambda^{12} & 2+\lambda^8 \\
      2-\lambda^{12} & -1+\lambda^8   & -1-\lambda^4 \\
      2+\lambda^8     & -1-\lambda^4    & 0
    \end{array}
  \right)
  +\frac{\lambda^{
      8-x_2}v_u^2}{M}
  \left(
    \begin{array}{ccc}
      \lambda^8 & \lambda^4 & -\lambda^4 \\
      \lambda^4 & 0         & 0          \\
      \lambda^4 & 0         & 0
    \end{array}
  \right). \label{neu-mass1}
\end{eqnarray}
We find that if the terms in the first and second lines are leading 
contributions, the tri-bimaximal mixing can be realized at the leading 
order. In order that the first term in the second line does not spoil 
the structures given in the first line, $m_1\ll m_2,~m_3$ is 
required. That is the reason why the neutrino mass spectrum in the 
cascade model should be the normal hierarchy. In the case, we can well 
approximated as 
\begin{eqnarray}
  m_2\simeq\sqrt{\Delta m_{21}^2}~~~\mbox{ and }~~~
  m_3\simeq\sqrt{|\Delta m_{31}^2|}.
\end{eqnarray}
Now we can obtain the following four constraints on the cascade 
parameters: (i) The neutrino masses should satisfy $m_1\ll m_2$. This 
means that $x_1\geq17$ for the parameters by utilizing~\eqref{m1} 
and~\eqref{m2}. This constraint leads to small mass of the lightest 
right-handed neutrino as shown later. (ii) In order to be consistent with a 
experimental data for the neutrino mass squared difference as 
\begin{eqnarray}
  r\equiv\frac{\sqrt{\Delta m_{21}^2}}{\sqrt{|\Delta m_{31}^2|}}\simeq0.18,
\end{eqnarray}
one should have a relation among the cascade parameters as $x_1-x_2=7$ 
or 8, where we use the fact that $\lambda\sim r$. (iii) We have a 
relation among the cascade parameters, light and heavy neutrino mass 
scales, that is, 
\begin{eqnarray}
  M\simeq\frac{\lambda^{
      8-x_2}v_u^2}{\sqrt{|\Delta m_{31}^2|}},
\end{eqnarray}
where $m_3\simeq\sqrt{|\Delta m_{31}^2|}$ is taken. (iv) The hierarchy 
$m_2\gg m_3\lambda^4$ is required in order that the second term in the 
last line of~\eqref{neu-mass1} does not spoil the democratic structure 
in the first line. This gives a constraint $x_1-x_2\geq5$. The above 
four constraints restrict the neutrino Dirac mass matrix of the 
cascade form and the right-handed one of the diagonal form to textures 
presented in Tab.~\ref{tab7} and~\ref{tab8}. We find that the minimal 
model for the neutrino mass matrices is described by 
$(x_1,x_2)=(17,10)$ given in Tab.~\ref{tab7}. In this case, 
mass spectrum of the right-handed neutrinos is estimated as
\begin{eqnarray}
  (M_1,M_2,M_3)\sim(10^5,10^
  {10},10^{16})\mbox{ GeV}.
\end{eqnarray}

Here we comment on the predicted mixing angles from cascade model. The 
mixing angles of the cascade model deviate from the exact 
tri-bimaximal mixing angles even if the~right-handed neutrino mass 
matrix is diagonal. The mixing angles can be estimated as 
\begin{eqnarray}
  \sin^2\theta_{12} &\simeq& \left|\frac{1}{\sqrt{3}}
    +\frac{2}{\sqrt{6}}\theta_{12}^{(1)}\right|^2
  \label{cor12} \\
  &\simeq& \left|\frac{1}{\sqrt{3}}
    -\frac{2}{\sqrt{3}}\frac{\bar{m}_1}{\bar{m}_2}
  \right|^2, \\
  \sin^2\theta_{23} &\simeq& \left|-\frac{1}{\sqrt{2}}
    -\frac{1}{\sqrt{6}}\theta_{13}^{(1)}
    +\frac{1}{\sqrt{3}}\theta_{23}^{(1)}\right|^2
  \label{cor23} \\
  &\simeq& \left|-\frac{1}{\sqrt{2}}
    +\frac{1}{\sqrt{2}}
    \frac{\bar{m}_1(3\bar{m}_3-\bar{m}_2)}
    {\bar{m}_3(\bar{m}_3-\bar{m}_2)}
    -\frac{\lambda^4}{3\sqrt{2}}
    \frac{\bar{m}_2}{\bar{m}_3-\bar{m}_2}\right|^2, \\
  \sin^2\theta_{13} &\simeq& \left|\frac{2}{\sqrt{6}}\theta_{13}^{(1)}
    +\frac{1}{\sqrt{3}}\theta_{23}^{(1)}\right|^2
  \label{cor13} \\
  &\simeq& \left|-\frac{\lambda^4}{\sqrt{2}}
    \frac{\bar{m}_3-\frac{2}{3}\bar{m}_2}
    {\bar{m}_3-\bar{m}_2}
    +\frac{\sqrt{2}\bar{m}_1\bar{m}_2}
    {\bar{m}_3(\bar{m}_3-\bar{m}_2)}\right|^2,
\end{eqnarray}
in a perturbative method,\footnote{See~\cite{su5} for a detailed 
  derivation.}  where parameters $\theta_{ij}^{(1)}$ indicate 
deviations from the exact tri-bimaximal mixing angles. These are 
elements of the following mixing matrix, 
\begin{eqnarray}
  V^{(1)}\simeq\left(
    \begin{array}{ccc}
      1 & \theta_{12}^{(1)} & \theta_{13}^{(1)} \\
      -\theta_{12}^{(1)} & 1 & \theta_{23}^{(1)} \\
      -\theta_{13}^{(1)} & -\theta_{23}^{(1)} & 1
    \end{array}
  \right). 
\end{eqnarray}
In our notation, the experimentally observed PMNS mixing matrix is 
given by $V_{\mbox{{\scriptsize PMNS}}}\simeq V_{\mbox{{\scriptsize 
      TB}}}V^{(1)}P_M$, where the $P_M$ is a diagonal phase matrix. 

\begin{table}
\begin{center}
\begin{tabular}{|c|c||c|c|}
\hline
$x_1$ & $x_2$ & $M_{\nu D}/
v_u
$ & $M_R/M$ \\
\hline
\hline
$17$ & $10$ &
$
\left(
\begin{array}{ccc}
\lambda^8 & \lambda^8 & \lambda^8 \\
\lambda^8 & \lambda^4  & -\lambda^4 \\
\lambda^8 & -\lambda^4 & 1
\end{array}
\right)
$ &
$
\left(
\begin{array}{ccc}
\lambda^{17} & 0           & 0 \\
0           & \lambda^{10} & 0 \\
0           & 0           & 1
\end{array}
\right)
$ \\
\hline
$18$ & $11$ &
$
\left(
\begin{array}{ccc}
\lambda^8 & \lambda^8  & \lambda^8  \\
\lambda^8 & \lambda^4  & -\lambda^4 \\
\lambda^8 & -\lambda^4 & 1
\end{array}
\right)
$ &
$
\left(
\begin{array}{ccc}
\lambda^{18} & 0           & 0 \\
0           & \lambda^{11} & 0 \\
0           & 0           & 1
\end{array}
\right)
$ \\
\hline
$\vdots$ & $\vdots$ & $\vdots$ & $\vdots$ \\
\hline
\end{tabular}
\end{center}
\caption{The textures of the neutrino Dirac mass matrix of the cascade 
form and the right-handed neutrino Majorana one of the diagonal 
from constrained by the experimentally observed values of the neutrino 
masses with the condition $x_1-x_2=7$.}
\label{tab7}
\end{table}
\begin{table}
\begin{center}
\begin{tabular}{|c|c||c|c|}
\hline
$x_1$ & $x_2$ & $M_{\nu D}/
v_u
$ & $M_R/M$ \\
\hline
\hline
$17$ & $9$ &
$
\left(
\begin{array}{ccc}
\lambda^8 & \lambda^8 & \lambda^8 \\
\lambda^8 & \lambda^4  & -\lambda^4 \\
\lambda^8 & -\lambda^4 & 1
\end{array}
\right)
$ &
$
\left(
\begin{array}{ccc}
\lambda^{17} & 0           & 0 \\
0           & \lambda^9 & 0 \\
0           & 0           & 1
\end{array}
\right)
$ \\
\hline
$18$ & $10$ &
$
\left(
\begin{array}{ccc}
\lambda^8 & \lambda^8  & \lambda^8  \\
\lambda^8 & \lambda^4  & -\lambda^4 \\
\lambda^8 & -\lambda^4 & 1
\end{array}
\right)
$ &
$
\left(
\begin{array}{ccc}
\lambda^{18} & 0           & 0 \\
0           & \lambda^{10} & 0 \\
0           & 0           & 1
\end{array}
\right)
$ \\
\hline
$\vdots$ & $\vdots$ & $\vdots$ & $\vdots$ \\
\hline
\end{tabular}
\end{center}
\caption{The textures of the neutrino Dirac mass matrix 
of the cascade form and the right-handed neutrino Majorana one 
of the diagonal from constrained by the experimentally observed 
values of the neutrino masses with the condition $x_1-x_2=8$.}
\label{tab8}
\end{table}

\subsubsection{Non-diagonal {\boldmath$M_R$} case}

We discuss the case of non-diagonal $M_R$, which is generically 
allowed in the context of the cascade textures. First, we define the 
diagonalized right-handed neutrino mass matrix, $D_R$, as 
\begin{eqnarray}
  D_R\equiv U_{\nu R}^TM_RU_{\nu R}\equiv
  \left(
    \begin{array}{ccc}
      \lambda^{x_1} & 0             & 0 \\
      0             & \lambda^{x_2} & 0 \\
      0             & 0             & 1
    \end{array}
  \right)M \hspace{3mm}\mbox{ with }\hspace{3mm}x_1\geq x_2\geq0,\label{DR}
\end{eqnarray}
where $M_R$ is a non-diagonal mass matrix for the right-handed 
neutrinos but mixing angles among each generation are assumed to be 
small in order to preserve the tri-bimaximal mixing. If the mixing 
angles among each generation of the right-handed neutrino are small 
enough, $U_{\nu R}$ can be written by 
\begin{eqnarray}
  U_{\nu R}\simeq
  \left(
    \begin{array}{ccc}
      1              & \theta_{R,12}  & \theta_{R,13} \\
      -\theta_{R,12} & 1              & \theta_{R,23} \\
      -\theta_{R,13} & -\theta_{R,23} & 1
    \end{array}
  \right)\equiv\left(
    \begin{array}{ccc}
      1                 & \lambda^{q_{12}}  & \lambda^{q_{13}} \\
      -\lambda^{q_{12}} & 1                 & \lambda^{q_{23}} \\
      -\lambda^{q_{13}} & -\lambda^{q_{23}} & 1
    \end{array}
  \right) \mbox{ with }q_{ij}\geq1,\label{UnuR}
\end{eqnarray}
up to the first order of $\theta_{R,ij}$ $(i,j=1\sim3)$. After the 
seesaw mechanism, we obtain the Majorana mass matrix of light 
neutrinos in low-energy as, 
\begin{eqnarray}
  M_\nu&\simeq&M_{\nu D}^TM_R^{-1}M_{\nu D}\nonumber\\
  &\simeq&\left[\lambda^{16}(M_R^{-1})_{11}\left(
      \begin{array}{ccc}
        1 & 1 & 1 \\
        1 & 1 & 1 \\
        1 & 1 & 1
      \end{array}\right)+\lambda^8(M_R^{-1})_{22}\left(
      \begin{array}{ccc}
        0 & 0  & 0  \\
        0 & 1  & -1 \\
        0 & -1 & 1
      \end{array}\right)\right.\nonumber\\
  &&\phantom{\Bigg[}+\lambda^8(M_R^{-1})_{22}\left(
    \begin{array}{ccc}
      \lambda^8 & \lambda^4 & \lambda^4 \\
      \lambda^4 & 0 & 0 \\
      \lambda^4 & 0 & 0
    \end{array}\right)\nonumber\\
  &&\phantom{\Bigg[}+(M_R^{-1})_{33}\left(
    \begin{array}{ccc}
      \lambda^{16}     & -\lambda^{12} & \lambda^8  \\
      -\lambda^{12} & \lambda^{12}  & -\lambda^4 \\
      \lambda^8      & -\lambda^4     & 1
    \end{array}\right)\nonumber\\
  &&\phantom{\Bigg[}+(M_R^{-1})_{23}\left(
    \begin{array}{ccc}
      2\lambda^{16} & 0 & \lambda^{8}(1-\lambda^4) \\
      0 & -2\lambda^8 & \lambda^4(1+\lambda^4) \\
      \lambda^8(1-\lambda^4) & \lambda^4(1+\lambda^4) &
      -2\lambda^4
    \end{array}\right)\nonumber\\
  &&\phantom{\Bigg[}+\lambda^8(M_R^{-1})_{12}\left(
    \begin{array}{ccc}
      2 & \lambda^8+\lambda^4 & \lambda^8-\lambda^4 \\
      \lambda^8+\lambda^4 & 2\lambda^4 & 0 \\
      \lambda^8-\lambda^4 & 0 & -2\lambda^4
    \end{array}\right)\nonumber\\
  &&\phantom{\Bigg[}\left.+\lambda^8(M_R^{-1})_{13}\left(
      \begin{array}{ccc}
        2\lambda^8 & \lambda^8-\lambda^4 & \lambda^8+1 \\
        \lambda^8-\lambda^4 & -2\lambda^4 & 1-\lambda^4 \\
        \lambda^8+1 & 1-\lambda^4 & 2
      \end{array}\right)\right]
  v_u^2.\label{neu-mass2}
\end{eqnarray}
When we operate the $V_{\mbox{{\scriptsize TB}}}$ to $M_\nu$ as 
$V_{\mbox{{\scriptsize TB}}}^TM_\nu V_{\mbox{{\scriptsize TB}}}$, the 
neutrino mass matrix is 
\begin{eqnarray}
  \mathcal{M}&\equiv&V_{\mbox{{\scriptsize TB}}}^TM_\nu
  V_{\mbox{{\scriptsize TB}}}\nonumber\\
  &\simeq&\left[3\lambda^{16}(M_R^{-1})_{11}\left(
      \begin{array}{ccc}
        0 & 0 & 0 \\
        0 & 1 & 0 \\
        0 & 0 & 0
      \end{array}\right)+2\lambda^8(M_R^{-1})_{22}\left(
      \begin{array}{ccc}
        0 & 0 & 0 \\
        0 & 0 & 0 \\
        0 & 0 & 1
      \end{array}\right)\right.\nonumber\\
  &&\phantom{\Bigg[}+\frac{\lambda^8(M_R^{-1})_{22}}{3}\left(
    \begin{array}{ccc}
      2\lambda^8 & \sqrt{2}\lambda^8 & -2\sqrt{3}\lambda^4 \\
      \sqrt{2}\lambda^8 & \lambda^8 & -\sqrt{6}\lambda^4 \\
      -2\sqrt{3}\lambda^4 & -\sqrt{6}\lambda^4 & 0
    \end{array}\right)\nonumber\\
  &&\phantom{\Bigg[}+\frac{(M_R^{-1})_{33}}{6}\left(
    \begin{array}{ccc}
      c_1^2           & -\sqrt{2}c_1c_2 & -\sqrt{3}c_1c_+ \\
      -\sqrt{2}c_1c_2 & 2c_2^2          & \sqrt{6}c_2c_+  \\
      -\sqrt{3}c_1c_+ & \sqrt{6}c_2c_+  & 3c_+^2
    \end{array}\right)\nonumber\\
  &&\phantom{\Bigg[}+\frac{(M_R^{-1})_{23}}{3\sqrt{2}}\left(
    \begin{array}{ccc}
      -2\sqrt{2}c_1\lambda^8 & c_3\lambda^8 &
      \sqrt{6}c_-(\lambda^4+\lambda^8) \\
      c_3\lambda^8 & 2\sqrt{2}c_2\lambda^8 &
      -\sqrt{3}c_-(2\lambda^4-\lambda^8) \\
      \sqrt{6}c_-(\lambda^4+\lambda^8) &
      -\sqrt{3}c_-(2\lambda^4-\lambda^8) & -6\sqrt{2}c_+\lambda^4
    \end{array}\right)\nonumber\\
  &&\phantom{\Bigg[}+\lambda^8(M_R^{-1})_{12}\left(
    \begin{array}{ccc}
      0 & \sqrt{2}\lambda^8 & 0 \\
      \sqrt{2}\lambda^8 & 2\lambda^8 & -\sqrt{6}\lambda^4 \\
      0 & -\sqrt{6}\lambda^4 & 0
    \end{array}\right)\nonumber\\
  &&\phantom{\Bigg[}
  \left.
    +\frac{\lambda^8(M_R^{-1})_{13}}{\sqrt{2}}\left(
      \begin{array}{ccc}
        0 & -c_1 & 0 \\
        -c_1 & 2\sqrt{2}c_2 & \sqrt{3}c_+ \\
        0 & \sqrt{3}c_+ & 0
      \end{array}\right)\right]
  v_u^2,
  \label{off-neu-mm}
\end{eqnarray}
where 
\begin{eqnarray}
  c_1\equiv1-\lambda^4-2\lambda^8,~~~
  c_2\equiv1-\lambda^4+\lambda^8,~~~
  c_3\equiv1-\lambda^4+4\lambda^8,~~~
  c_\pm\equiv1\pm\lambda^4.
\end{eqnarray}
This mass matrix can be rewritten by
\begin{eqnarray}
  \label{mroffmas}
  \mathcal{M}=\mathcal{M}_0+\mathcal{M}_{\mbox{{\scriptsize off}}}
  \equiv\mathcal{M}_0+\left(
    \begin{array}{ccc}
      m_1^R & m_{12}^R & m_{13}^R \\
      m_{12}^R & m_2^R & m_{23}^R \\
      m_{13}^R & m_{23}^R & m_3^R
    \end{array}
  \right),
\end{eqnarray}
where $\mathcal{M}_0$ comes from the diagonal elements of $M_R$, which 
is given by 
\begin{eqnarray}
  \mathcal{M}_0\simeq\left(
    \begin{array}{ccc}
      \bar{m}_1+\frac{\lambda^8}{3}\bar{m}_3 &
      -\sqrt{2}\bar{m}_1+\frac{\sqrt{2}\lambda^4}{6}\bar{m}_3 &
      -\sqrt{3}\bar{m}_1-\frac{\lambda^4}{\sqrt{3}}\bar{m}_3 \\
      -\sqrt{2}\bar{m}_1+\frac{\sqrt{2}\lambda^8}{6}\bar{m}_3 &
      \bar{m}_2+2\bar{m}_1+\frac{\lambda^8}{6}\bar{m}_3 &
      \sqrt{6}\bar{m}_1-\frac{\lambda^4}{\sqrt{6}}\bar{m}_3 \\
      -\sqrt{3}\bar{m}_1-\frac{\lambda^4}{\sqrt{3}}\bar{m}_3 &
      \sqrt{6}\bar{m}_1-\frac{\lambda^4}{\sqrt{6}}\bar{m}_3 &
      \bar{m}_3+3\bar{m}_1
    \end{array}
  \right).
  \label{neu-mass-diag1}
\end{eqnarray}
In~\eqref{mroffmas} $\mathcal{M}_{\mbox{{\scriptsize off}}}$ has
effects from the off-diagonal elements of $M_R$. In order to obtain 
experimentally accepted mixing structure without unnatural 
cancellations, we focus only on a case that the collections from the off-diagonal elements of 
$M_R$ are small enough not to spoil the nearly tri-bimaximal mixing 
constructed by the cascade neutrino Dirac mass matrix. This means that 
the resultant structure of neutrino mass matrix given 
in~\eqref{off-neu-mm} should not be drastically differed from 
the~\eqref{neu-mass-diag1}, and thus the magnitude of neutrino mass 
eigenvalues~\eqref{m1}--\eqref{m3} and the above four constraints 
should be satisfied at the leading order even in non-diagonal $M_R$ 
case. These discussions give the following neutrino mass eigenvalues 
up to the next leading order,\footnote{Detailed discussions is given 
  in the Appendix.} 
\begin{eqnarray}
  m_1 &\simeq& \frac{
    v_u^2}{6M}+m_1^R=\bar{m_1}+m_1^R, \\
  m_2 &\simeq& \left(3\lambda^{16-x_1}+\frac{1}{3}\right)
  \frac{
    v_u^2}{M}+m_2^R=\bar{m}_2+m_2^R+2\bar{m_1}, \\
  m_3 &\simeq& \left(2\lambda^{8-x_2}+\frac{1}{2}\right)
  \frac{
    v_u^2}{M}+m_3^R=\bar{m_3}+m_3^R+3\bar{m_3},
\end{eqnarray}
where $m_i^R$ include effects from the off-diagonal element of $M_R$ 
described by 
\begin{eqnarray}
  m_1^R &\equiv& \frac{
    v_u^2}{6M}\lambda^{-x_1}\theta_{R,23}^2, \\
  m_2^R &\equiv& \frac{
    v_u^2}{M}
  (\lambda^{-x_2}\theta_{R,23}^2-2\lambda^{8-x_1}\theta_{R,13}),
  \\
  m_3^R &\equiv& \frac{
    v_u^2}{2M}\lambda^{-x_1}
  (2\lambda^4\theta_{R,12}-\theta_{R,13})^2.
\end{eqnarray}
Typical textures of non-diagonal $M_R$ are presented in 
Tabs.~\ref{tab9} and~\ref{tab10}. 
\begin{table}
\begin{center}
\begin{tabular}{|c|c||c|c|}
\hline
$x_1$ & $x_2$ & $M_{\nu D}/
v_u
$ & $M_R/M$ \\
\hline
\hline
$17$ & $10$ &
$
\left(
\begin{array}{ccc}
\lambda^8 & \lambda^8 & \lambda^8 \\
\lambda^8 & \lambda^4  & -\lambda^4 \\
\lambda^8 & -\lambda^4 & 1
\end{array}
\right)
$ &
$
\left(
\begin{array}{ccc}
\lambda^{17} & \lambda^{17}           & \lambda^{10} \\
\lambda^{17}           & \lambda^{10} & \lambda^7 \\
\lambda^{10}           & \lambda^7           & 1
\end{array}
\right)
$ \\
\hline
$18$ & $11$ &
$
\left(
\begin{array}{ccc}
\lambda^8 & \lambda^8  & \lambda^8  \\
\lambda^8 & \lambda^4  & -\lambda^4 \\
\lambda^8 & -\lambda^4 & 1
\end{array}
\right)
$ &
$
\left(
\begin{array}{ccc}
\lambda^{18} & \lambda^{19}           & \lambda^{11} \\
\lambda^{19}           & \lambda^{11} & \lambda^8 \\
\lambda^{11}           & \lambda^8           & 1
\end{array}
\right)
$ \\
\hline
$\vdots$ & $\vdots$ & $\vdots$ & $\vdots$ \\
\hline
\end{tabular}
\end{center}
\caption{The textures of the neutrino Dirac mass matrix of the cascade
form and the right-handed neutrino Majorana one of the non-diagonal from. 
The matrices are constrained by the experimentally observed values of 
the neutrino masses with the condition $x_1-x_2=7$.}
\label{tab9}
\end{table}
\begin{table}
\begin{center}
\begin{tabular}{|c|c||c|c|}
\hline
$x_1$ & $x_2$ & $M_{\nu D}/
v_u
$ & $M_R/M$ \\
\hline
\hline
$17$ & $9$ &
$
\left(
\begin{array}{ccc}
\lambda^8 & \lambda^8 & \lambda^8 \\
\lambda^8 & \lambda^4  & -\lambda^4 \\
\lambda^8 & -\lambda^4 & 1
\end{array}
\right)
$ &
$
\left(
\begin{array}{ccc}
\lambda^{17} & \lambda^{16}           & \lambda^{10} \\
\lambda^{16}           & \lambda^9 & \lambda^6 \\
\lambda^{10}           & \lambda^6           & 1
\end{array}
\right)
$ \\
\hline
$18$ & $10$ &
$
\left(
\begin{array}{ccc}
\lambda^8 & \lambda^8  & \lambda^8  \\
\lambda^8 & \lambda^4  & -\lambda^4 \\
\lambda^8 & -\lambda^4 & 1
\end{array}
\right)
$ &
$
\left(
\begin{array}{ccc}
\lambda^{18} & \lambda^{18}           & \lambda^{10} \\
\lambda^{18}           & \lambda^{10} & \lambda^7 \\
\lambda^{10}           & \lambda^7           & 1
\end{array}
\right)
$ \\
\hline
$\vdots$ & $\vdots$ & $\vdots$ & $\vdots$ \\
\hline
\end{tabular}
\end{center}
\caption{The textures of the neutrino Dirac mass matrix of the cascade 
form and the right-handed neutrino Majorana one of the diagonal from. 
The matrices constrained by the  experimentally observed values of 
the neutrino masses with the condition $x_1-x_2=8$.}
\label{tab10}
\end{table}
The collections to the generation mixing angles are also estimated as
\begin{eqnarray}
  \theta_{12}^{(1)}&\simeq&-\frac{\sqrt{2}\bar{m}_1+m_{12}^R}{\bar{m}_2+m_2^R},
  \\
  \theta_{23}^{(1)}&\simeq&\frac{\sqrt{6}\bar{m}_1-\frac{\lambda^4}{\sqrt{6}}\bar{m}_3+m_{23}^R}{(\bar{m}_3+m_3^R)-(\bar{m}_2+m_2^R)},
  \\
  \theta_{13}^{(1)}&\simeq&\frac{-\sqrt{3}\bar{m}_1-\frac{\lambda^4}{\sqrt{3}}\bar{m}_3+m_{13}^R}{\bar{m}_3+m_3^R},
\end{eqnarray}
where
\begin{eqnarray}
  m_{12}^R &\simeq& -\frac{1}{6\sqrt{2}}
  [\lambda^{-8}\theta_{R,23}^2\bar{m}_3
  -2(2\theta_{R,12}-\lambda^{-8}\theta_{R,13})\bar{m}_2],
  \\
  m_{23}^R &\simeq&
  \frac{1}{\sqrt{6}}[\theta_{R,23}\bar{m}_3+\lambda^{-8}(2\lambda^{d_2}\theta_{R,12}-\theta_{R,13})(1-\lambda^{-8}\theta_{R,13})\bar{m}_2],
  \\
  m_{13}^R &\simeq& -\frac{1}{2\sqrt{3}}
  \left[\frac{\lambda^{-4}}{2}(2+\lambda^{-4}\theta_{R,23})
    \theta_{R,23}\bar{m}_3
    -\frac{4}{3}\lambda^{-4}\theta_{R,12}^2\bar{m}_2
  \right].
\end{eqnarray}
Finally, the PMNS mixing angles including collections from the 
off-diagonal elements are 
\begin{eqnarray}
  \sin\theta_{12}
  &\simeq& \frac{1}{\sqrt{3}}
  +\frac{2}{\sqrt{6}}\frac{-\bar{m}_1+m_{12}^R}{\bar{m}_2+m_2^R},
  \\
  \sin\theta_{23}
  &\simeq&-\frac{1}{\sqrt{2}}
  +\frac{1}{\sqrt{2}}
  \frac{\bar{m}_1[3(\bar{m}_3+m_3^R)-(\bar{m}_2+m_2^R)]}
  {(\bar{m}_3+m_3^R)[(\bar{m}_3+m_3^R)-(\bar{m}_2+m_2^R)]}
  \nonumber \\
  &      & -\frac{\lambda^{d_1-d_2}}{3\sqrt{2}}
  \frac{\bar{m}_3(\bar{m}_2+m_2^R)}
  {(\bar{m}_3+m_3^R)[(\bar{m}_3+m_3^R)-(\bar{m}_2+m_2^R)}
  -\frac{1}{\sqrt{6}}\frac{m_{13}^R}{\bar{m}_3+m_3^R} \nonumber \\
  &      & +\frac{1}{\sqrt{3}}
  \frac{m_{23}^R}{(\bar{m}_3+m_3^R)-(\bar{m}_2+m_2^R)},\\
  \sin\theta_{13}
  &\simeq& -\frac{\lambda^{d_1-d_2}}{\sqrt{2}}
  \frac{\bar{m}_3\left[(\bar{m}_3+m_3^R)
      -\frac{2}{3}(\bar{m}_2+m_2^R)\right]}
  {(\bar{m}_3+m_3^R)[(\bar{m}_3+m_3^R)-(\bar{m}_2+m_2^R)]}
  \nonumber \\
  &     & +\frac{\sqrt{2}\bar{m}_1(\bar{m}_2+m_2^R)}
  {(\bar{m}_3+m_3^R)[(\bar{m}_3+m_3^R)-(\bar{m}_2+m_2^R)]}
  +\frac{2}{\sqrt{6}}\frac{m_{13}^R}{\bar{m}_3+m_3^R} \nonumber \\
  &     & +\frac{1}{\sqrt{3}}
  \frac{m_{23}^R}{(\bar{m}_3+m_3^R)-(\bar{m}_2+m_2^R)}.
\end{eqnarray}

\subsection{Charged lepton and quark sectors}

At the end of this section, we study the charged lepton and quark 
sectors. Under the condition of the $SO(10)$ scenario, we examine 
quantitative features of the masses and mixing angles. 

We take the charged lepton mass matrix as 
\begin{eqnarray}
  M_e\simeq
  \left(
    \begin{array}{ccc}
      \epsilon_d & \delta_d    & \delta_d \\
      \delta_d   & -3\lambda_d & -3\lambda_d \\
      \delta_d   & -3\lambda_d   & 1
    \end{array}
  \right)\xi_dv_d,
\end{eqnarray}
in our study. The magnitudes of cascade parameters can be partially 
evaluated from the observed values of charged lepton masses, and are 
given by $|\lambda_d|\simeq m_\mu/(3m_\tau)$ and 
$|\delta_d|\simeq3\sqrt{m_em_\mu}/m_\tau$. We find that the 
corrections from the charged lepton sector are generally small; the 
total leptonic mixing angles can be written as 
\begin{eqnarray}
  \sin\theta_{12}
  &\simeq& \frac{1}{\sqrt{3}}
  +\frac{2}{\sqrt{6}}\frac{-\bar{m}_1+m_{12}^R}{\bar{m}_2+m_2^R}
  +\sqrt{\frac{3m_e}{m_\mu}}, \\
  \sin\theta_{23}
  &\simeq&-\frac{1}{\sqrt{2}}
  +\frac{1}{\sqrt{2}}
  \frac{\bar{m}_1[3(\bar{m}_3+m_3^R)-(\bar{m}_2+m_2^R)]}
  {(\bar{m}_3+m_3^R)[(\bar{m}_3+m_3^R)-(\bar{m}_2+m_2^R)]}
  \nonumber \\
  &      & -\frac{\lambda^{d_1-d_2}}{3\sqrt{2}}
  \frac{\bar{m}_3(\bar{m}_2+m_2^R)}
  {(\bar{m}_3+m_3^R)[(\bar{m}_3+m_3^R)-(\bar{m}_2+m_2^R)}
  -\frac{1}{\sqrt{6}}\frac{m_{13}^R}{\bar{m}_3+m_3^R} \nonumber \\
  &      & +\frac{1}{\sqrt{3}}
  \frac{m_{23}^R}{(\bar{m}_3+m_3^R)-(\bar{m}_2+m_2^R)}
  +\frac{m_\mu}{3m_\tau},\\
  \sin\theta_{13}
  &\simeq& -\frac{\lambda^{d_1-d_2}}{\sqrt{2}}
  \frac{\bar{m}_3\left[(\bar{m}_3+m_3^R)
      -\frac{2}{3}(\bar{m}_2+m_2^R)\right]}
  {(\bar{m}_3+m_3^R)[(\bar{m}_3+m_3^R)-(\bar{m}_2+m_2^R)]}
  \nonumber \\
  &     & +\frac{\sqrt{2}\bar{m}_1(\bar{m}_2+m_2^R)}
  {(\bar{m}_3+m_3^R)[(\bar{m}_3+m_3^R)-(\bar{m}_2+m_2^R)]}
  +\frac{2}{\sqrt{6}}\frac{m_{13}^R}{\bar{m}_3+m_3^R} \nonumber \\
  &     & +\frac{1}{\sqrt{3}}
  \frac{m_{23}^R}{(\bar{m}_3+m_3^R)-(\bar{m}_2+m_2^R)}
  +\frac{3}{\sqrt{2}}\sqrt{\frac{m_e}{m_\mu}},
\end{eqnarray}
at the first order of perturbations. 

Next, we comment on the quark sector. One must remember that the mass 
matrix of the H.C. form is motivated for the mass spectra and mixing 
angles of quark sector. The mixing matrices for the up- and down-sector are 
given by the cascading mass matrices \eqref{Mu-cas} and~\eqref{Md-cas}. From 
the mass matrices, one can estimate the following mixing angles
\begin{eqnarray}
  V_d =
  \left(
    \begin{array}{lll}
      \mathcal{O}(1) & \mathcal{O}(\lambda) & \mathcal{O}(\lambda^3) \\
      \mathcal{O}(\lambda) & \mathcal{O}(1) & \mathcal{O}(\lambda^2) \\
      \mathcal{O}(\lambda^3) & \mathcal{O}(\lambda^2) & \mathcal{O}(1)
    \end{array}
  \right),~~~
  V_u =
  \left(
    \begin{array}{lll}
      \mathcal{O}(1) & \mathcal{O}(\lambda^4) & \mathcal{O}(\lambda^8) \\
      \mathcal{O}(\lambda^4) & \mathcal{O}(1) & \mathcal{O}(\lambda^4) \\
      \mathcal{O}(\lambda^8) & \mathcal{O}(\lambda^4) & \mathcal{O}(1)
    \end{array}
  \right), 
\end{eqnarray}
where $V_d$ and $V_u$ are unitary mixing matrices determined by $M_u$ 
and $M_d$. It can be easily seen from the structure of $V_d$ that the 
experimentally observed values of CKM matrix can be realized at the 
leading order and the collections from the $V_u$ are generally 
small. Detailed numerical calculations are given in the next section. 

\section{Phenomenologies}

In this section, we perform the numerical study of phenomenologies 
based on the above analyses of the cascade textures for the quark and 
lepton sectors: the PMNS mixing angles, lepton flavor violation (LFV), 
baryon asymmetry of the Universe (BAU) via thermal leptogenesis. 

\subsection{PMNS mixing angles}

Firstly, we show numerical analyses of the generation mixing angles of 
the quark and lepton sectors predicted from the cascade model. Here, we 
investigate two typical types of minimal texture for the neutrino Dirac and 
right-handed Majorana neutrino mass matrices,
\begin{eqnarray}
  \mbox{Model I : }~~~
  M_{\nu D} &=&
  \left(
    \begin{array}{ccc}
      c_\nu\lambda^8 & c_\nu\lambda^8  & c_\nu\lambda^8 \\
      c_\nu\lambda^8 & b_\nu\lambda^4  & -b_\nu\lambda^4 \\
      c_\nu\lambda^8 & -b_\nu\lambda^4 & a_\nu
    \end{array}
  \right)
  v_u, \\
  M_R &=&
  \left(
    \begin{array}{ccc}
      f_R\lambda^{17} & e_R\lambda^{17} & d_R\lambda^{10} \\
      e_R\lambda^{17} & c_R\lambda^{10} & b_R\lambda^7 \\
      d_R\lambda^{10} & b_R\lambda^7    & a_R
    \end{array}
  \right)M,
\end{eqnarray}
and 
\begin{eqnarray}
  \mbox{Model II : }~~~
  M_{\nu D} &
  =&
  \left(
    \begin{array}{ccc}
      c_\nu\lambda^8 & c_\nu\lambda^8  & c_\nu\lambda^8 \\
      c_\nu\lambda^8 & b_\nu\lambda^4  & -b_\nu\lambda^4 \\
      c_\nu\lambda^8 & -b_\nu\lambda^4 & a_\nu
    \end{array}
  \right)
  v_u, \\
  M_R &
  =&
  \left(
    \begin{array}{ccc}
      f_R\lambda^{17} & e_R\lambda^{16} & d_R\lambda^{10} \\
      e_R\lambda^{16} & c_R\lambda^{10} & b_R\lambda^6 \\
      d_R\lambda^{10} & b_R\lambda^6    & a_R
    \end{array}
  \right)M,
\end{eqnarray}
for the cases of the condition $x_1-x_2=7$.\footnote{We focus on only the 
$x_1-x_2=7$ case given in Tab. \ref{tab9} as a typical example (Model I). In 
order to see effects from off-diagonal elements of right-handed neutrino mass 
matrix to the PMNS mixing angles, we also analyse a slightly different model 
for $M_R$ as a comparison (Model II).} Here $a_\nu$, $b_\nu$ $c_\nu$, and 
$a_R$, $\cdots$, $f_R$ are complex numbers whose absolute values are taken as 
$0.4\sim1.4$. In both models, the following charged lepton, up and down quark 
mass matrices are utilized 
\begin{eqnarray}
  M_e=
  \left(
    \begin{array}{ccc}
      0            & e_e\lambda^3   & d_e\lambda^3 \\
      e_e\lambda^3 & -3c_e\lambda^2 & -3b_e\lambda^2 \\
      d_e\lambda^3 & -3b_e\lambda^2   & a_e
    \end{array}
  \right)\lambda v_d,
\end{eqnarray}
and
\begin{eqnarray}
  M_u=
  \left(
    \begin{array}{ccc}
      c_\nu\lambda^8 & c_\nu\lambda^8  & c_\nu\lambda^8 \\
      c_\nu\lambda^8 & b_\nu\lambda^4  & -b_\nu\lambda^4 \\
      c_\nu\lambda^8 & -b_\nu\lambda^4 & a_\nu
    \end{array}
  \right)v_u,~~~
  M_d=
  \left(
    \begin{array}{ccc}
      0            & e_e\lambda^3   & d_e\lambda^3 \\
      e_e\lambda^3 & c_e\lambda^2 & b_e\lambda^2 \\
      d_e\lambda^3 & b_e\lambda^2   & a_e
    \end{array}
  \right)\lambda v_d,
\end{eqnarray}
where $a_e$, $\cdots$, $e_e$ are also complex values whose range are 
the same as $a_\nu$, $b_\nu$, $c_\nu$, and $a_R$, $\cdots$, $f_R$. 
Moreover, notice that the mass matrices of down and up quarks have the 
same structures of ones of charged lepton, except for the GJ factor, 
and neutrino respectively because of the mass relations of $SO(10)$ 
model at the GUT scale. 

\begin{figure}
\begin{center}
Model I\vspace{2mm}

\includegraphics[scale = 0.9]{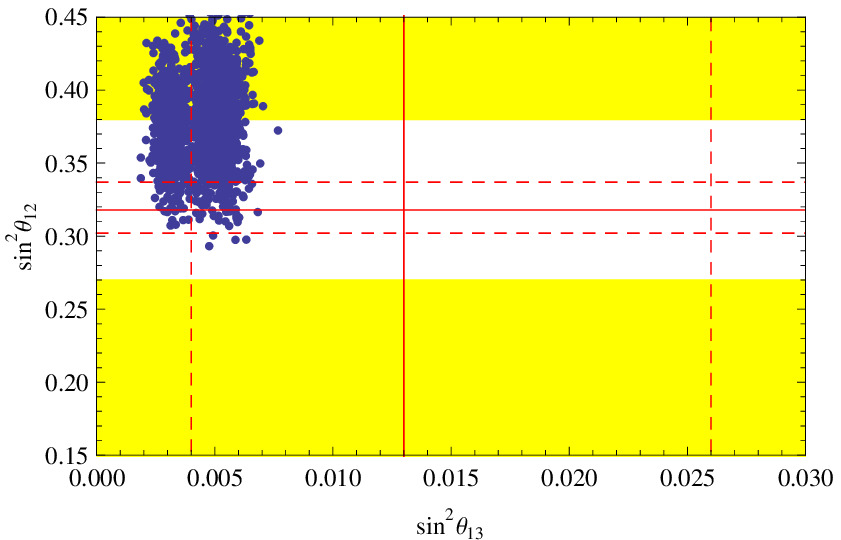}
\includegraphics[scale = 0.9]{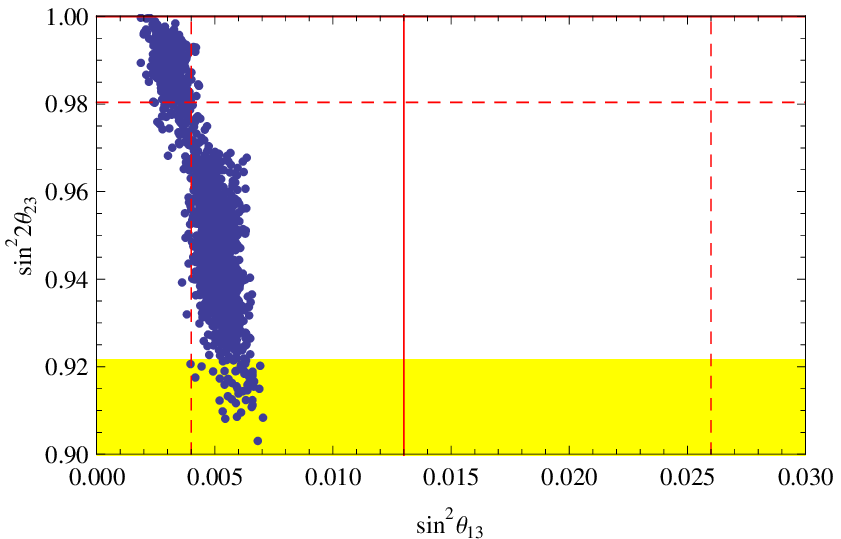}

Model II\vspace{2mm}

\includegraphics[scale = 0.9]{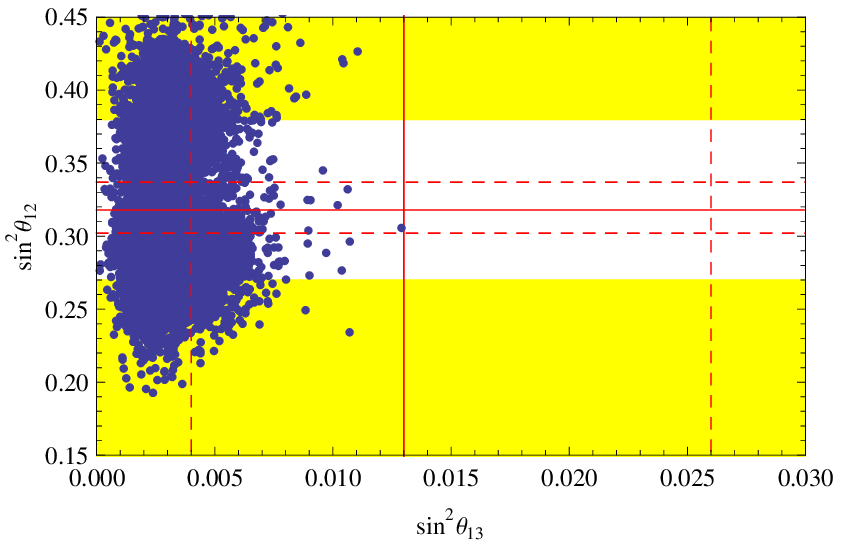}
\includegraphics[scale = 0.9]{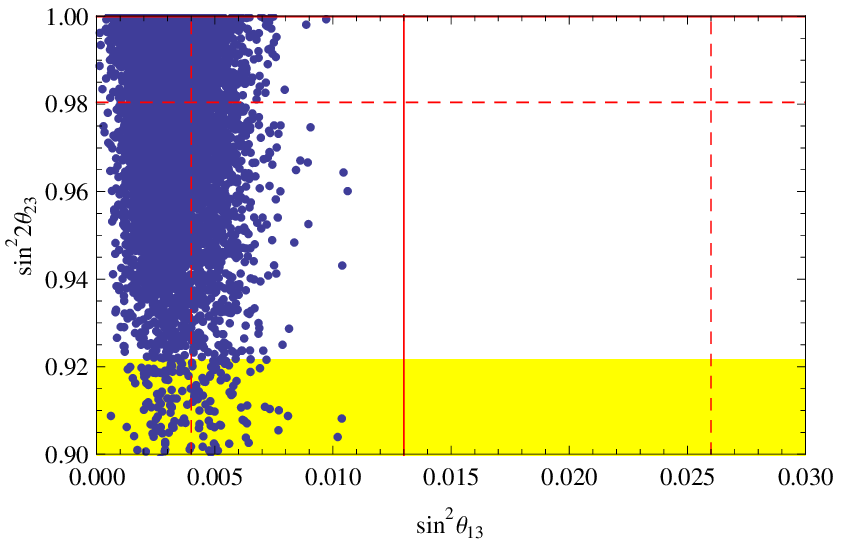}
\end{center}
\caption{Predicted PMNS mixing angles from Model I and II: The solid and
dashed lines, and white background correspond to the best fit value, 
boundaries of $1\sigma$ errors and the region within $3\sigma$.}
\label{fig1}
\end{figure}

The results of numerical calculation for the PMNS mixing angles in 
Model I and II are given in Figs.~\ref{fig1}. 
The all plots can fit the data of quark masses, CKM mixing angles, 
charged lepton masses, and the mass ratio of two mass squared 
differences of neutrino. These are 
\begin{eqnarray}
  &&
  \frac{m_u}{m_c}=0.0026(6),~~~\frac{m_c}{m_t}=0.0023(2),~~~y_t=0.51(2),
  \\
  && \frac{m_d}{m_s}=0.051(7),~~~\frac{m_s}{m_b}=0.018(2),~~~y_b=0.34(3), \\
  && A=0.73(3),~~~\lambda_c=0.227(1),~~~\bar{\rho}=0.22(6),~~~
  \bar{\eta}=0.33(4),
\end{eqnarray}
at the GUT scale, where the numbers in parentheses mean an uncertainty 
in last digit~\cite{Ross:2007az}.\footnote{We have taken $a_\nu$ and 
  $a_e$ as $0.49\leq a_\nu\leq0.53$ and $0.31\leq a_e\lambda\leq 0.37$ 
  to fit $y_t$ and $y_b$ in our numerical analyses since the values of 
  $y_t$ and $y_b$ are almost determined by ones of $a_\nu$ and $a_e$, 
  respectively.} Here $A$, $\lambda_c$, $\bar{\rho}$, and $\bar{\eta}$ 
are the Wolfenstein parameters which correspond to three mixing angles 
and one phase in the CKM matrix; $y_{t}$ and $y_b$ represent largest 
eigenvalues of Yukawa matrices for the up-type and down-type quarks, 
respectively. These values are calculated with the 2-loop gauge 
coupling and 2-loop Yukawa coupling renormalization equation taking 
$\tan\beta$, threshold corrections $\gamma_{t,b,d}$, and an effective 
SUSY scale as $\tan\beta=38$, $\gamma_b=-0.22$, $\gamma_d=-0.21$,
$\gamma_t=0$, and $\msusy=500$ GeV, respectively. The threshold 
corrections are approximated by
\begin{eqnarray}
  \gamma_t\sim y_t^2\frac{\tan\beta}{32\pi^2}\frac{\mu A_t}{m_{\tilde{t}}^2},
  ~~~
  \gamma_u\sim 0,~~~
  \gamma_b\sim\frac{4}{3}g_3^2\frac{\tan\beta}{16\pi^2}
  \frac{\mu \bar{M}_3}{m_{\tilde{b}}^2},~~~
  \gamma_d\sim\frac{4}{3}g_3^2\frac{\tan\beta}{16\pi^2}
  \frac{\mu \bar{M}_3}{m_{\tilde{d}}^2},
\end{eqnarray}
where $\mu$, $A_t$, $m_{\tilde{q}}$, $g_3$, and $M_3$ indicate 
supersymmetric Higgs mass $\mu H_uH_d$, soft top quark tri-linear 
coupling, mass of the squark $\tilde{q}$, strong coupling, and gaugino 
soft breaking mass, respectively~\cite{Carena:1999py}. The utilized 
values of SUSY can lead to the relations at the GUT scale, 
\begin{eqnarray}
  \frac{m_b}{m_\tau}=\frac{3m_s}{m_\mu}=\frac{m_d}{3m_e}=1,
\end{eqnarray}
in a good accuracy. Therefore, the above values can automatically 
reproduce the experimental observed charged lepton masses at the low 
energy. 

The numerical calculations of Model I suggest that the predicted
region of solar angle covers the experimental upper bound but the
model gives a predicted lower bound around
$0.29\lesssim\sin^2\theta_{12}$. On the other hand, a constrained
region is predicted for the reactor angle as
$0.002\lesssim\sin^2\theta_{13}\lesssim0.007$. This result is one of
important predictions of the present cascade textures. It might be
checked by the upcoming DoubleChooz~\cite{Ardellier:2006mn},
RENO~\cite{:2010vy}, and DayaBay~\cite{Guo:2007ug} experiments as the
reactor experiments in addition to the accelerator experiments such as
T2K~\cite{Itow:2001ee} and
NO$\nu$A~\cite{Ayres:2004js}.\footnote{See~\cite{Mezzetto:2010zi} for
  an excellent review of sensitivities of the upcoming experiments.}
Finally, the atmospheric angle covers the current experimentally
allowed region. However, there is a relatively clear correlation
between the magnitudes of reactor and atmospheric angles in
Fig.~\ref{fig1}. On the other hand, the result for Model II can
cover the experimental allowed region because of largeness of
corrections from the right-handed neutrino mass matrix but there is an
upper bound of $\sin^2\theta_{13}$ which is
$\sin^2\theta_{13}\lesssim0.015$. Therefore, we can conclude that the
minimal cascade textures in the context of SUSY $SO(10)$ (Model I) can
lead to clear predictions, which are $0.29\lesssim\sin^2\theta_{12}$
and $0.002\lesssim\sin^2\theta_{13}\lesssim0.007$, and relatively
sharp correlations between the reactor and atmospheric angles. In the
next-to minimal cascade model (Model II) lead to only the upper bound
of the reactor angle while the model can explain about almost ranges
of PMNS mixing angles. The minimal realization of cascade model is
predictive and interesting in the framework with the cascade
hierarchies in SUSY $SO(10)$ GUT, and thus it might be checked by the
future experiments. It would be too difficult to distinguish other
cascade textures even if the future precision data of neutrino
oscillation experiment could be used. It is worth studying 
a new method to check the models.

At the end of this subsection, we give a brief comparison between our 
results and ones from a similar $SO(10)$ approach, which utilizes type II 
seesaw mechanism and a simple ansatz such that the dominant Yukawa matrix has 
rank one \cite{Dutta:2009ij}.\footnote{The paper by Dutta, et al. in refs. 
\cite{Altarelli:2005yp} presents an $S_4$ flavor model, which is one of 
realizations of the rank I approach by flavor symmetry. Here we focus on the 
general results of this approach given in \cite{Dutta:2009ij}.} The work gives 
some predicted regions for leptonic mixing angles based on three typical models
 in this approach: (A) $V_\nu=1$ case where 
$V_{\text{PMNS}}=V_eV_{\text{TB}}V_\nu^\dagger$, and $V_e$ and $V_\nu$ are 
diagonalizing matrices of Yukawa matrix for the charged lepton and neutrino 
mass matrix for light neutrinos in a tetrahedral coordinate, respectively, (B1)
 $V_\nu\neq1$ and 
$(f^{\text{tetra}})_{12}=(f^{\text{tetra}})_{21}=(f^{\text{tetra}})_{13}=(f^{\text{tetra}})_{31}=0$ where $f^{\text{tetra}}$ is a coupling to $126+\bar{126}$ 
Higgs in tetrahedral coodinate, and (B2) $V_\nu\neq1$ and 
$(f^{\text{tetra}})_{12}=(f^{\text{tetra}})_{21}=(f^{\text{tetra}})_{23}=(f^{\text{tetra}})_{32}=0$. The numerical calculations predict 
$\sin^2\theta_{12}\simeq 0.28$ and 
$0.006\lesssim\sin^2\theta_{13}\lesssim0.008$ in the model A, 
$0.32\lesssim\sin^2\theta_{12}\lesssim0.33$, 
$0.003\lesssim\sin^2\theta_{13}\lesssim0.006$, and 
$0.59\lesssim\sin^2\theta_{23}\lesssim0.60$ in the model B1, and 
$\sin^2\theta_{12}\simeq 0.30$, $0.04\lesssim\sin^2\theta_{13}\lesssim0.06$, 
and $0.62\lesssim\sin^2\theta_{23}\lesssim0.63$ in the model B2 by using the 
data $\Delta m_{21}^2/\Delta m_{31}^2=0.027-0.038$ at $2\sigma$ level. These can be compared with our results and might be also checked by the 
future neutrino experiments.

\subsection{Lepton flavor violation}

Next, we investigate the branching ratios of LFV process 
$l_i\rightarrow l_j\gamma$, in our cascade lepton mass matrices. We 
suppose that soft SUSY breaking masses of sleptons are universal at 
the GUT scale, $\LamGUT$, for simplicity. In the case, the 
off-diagonal matrix elements are generated by radiative corrections 
from the Yukawa couplings of neutrinos~\cite{Borzumati:1986qx}. The 
one-loop renormalization group evolution gives the left-handed slepton 
masses. The leading contribution is estimated by 
\begin{eqnarray}
  (m_l^2)_{ij}\sim\frac{3m_0^2+|a_0|^2}{8\pi^2v^2\sin^2\beta}
  \sum_k(M_{\nu D}^\dagger)_{ik}
  (M_{\nu D})_{kj}\ln\left(\frac{|M_k|}{\LamGUT}\right)~~~
  (\mbox{for }i\neq j),
\end{eqnarray}
where $m_0$ and $a_0$ are the universal SUSY breaking mass and 
three-point coupling of scalar superpartners given at the GUT 
scale. The branching fractions of each LFV process are roughly given 
by
\begin{eqnarray}
  \mbox{Br}(l_i\rightarrow l_j\gamma)
  \simeq\frac{3\alpha}{2\pi}\frac{|(m_l^2)_{ij}|^2M_W^4}{\msusy^8}\tan^2\beta
  ,
\end{eqnarray}
where $\alpha$, $M_W$, and $\msusy$ are the fine structure constant, 
$W$ boson mass, and a typical mass scale of superparticles, 
respectively. These branching ratios are estimated by 
\begin{eqnarray}
  \mbox{Br}(\mu\rightarrow e\gamma)
  &\simeq& \frac{3\alpha}{8\pi^5}
  B
  \left[\lambda^{16}\ln\left(\frac{|M_1|}{\LamGUT}\right)
    +\lambda^{12}\ln\left(\frac{|M_2|}{\LamGUT}\right)
    -\lambda^{12}\ln\left(\frac{|M_3|}{\LamGUT}\right)\right]^2,
  \nonumber \\ && \\
  \mbox{Br}(\tau\rightarrow e\gamma)
  &\simeq& \frac{3\alpha}{8\pi^5}
  B
  \left[\lambda^{16}\ln\left(\frac{|M_1|}{\LamGUT}\right)
    -\lambda^{12}\ln\left(\frac{|M_2|}{\LamGUT}\right)
    +\lambda^8\ln\left(\frac{|M_3|}{\LamGUT}\right)\right]^2,
  \nonumber \\ && \\
  \mbox{Br}(\tau\rightarrow\mu\gamma)
  &\simeq& \frac{3\alpha}{8\pi^5}
  B
  \left[\lambda^{12}\ln\left(\frac{|M_1|}{\LamGUT}\right)
    -\lambda^8\ln\left(\frac{|M_2|}{\LamGUT}\right)
    -\lambda^4\ln\left(\frac{|M_3|}{\LamGUT}\right)\right]^2,
\end{eqnarray}
for both Model I and II. Here we define $B\equiv(M_W/\msusy)^4\tan^2\beta$ and 
take $m_0=|a_0|=\msusy$. Typical magnitudes of the branching ratios are shown 
in Tab.~\ref{tab11}. 
\begin{table}
\begin{center}
\begin{tabular}{
|c|c|c|c|c|c|}
\hline
$\mbox{Br}(\mu\rightarrow e\gamma)/B$ & $\mbox{Br}(\tau\rightarrow
e\gamma)/B$ & $\mbox{Br}(\tau\rightarrow\mu\gamma)/B$ & $M_1$ [GeV] &
$M_2$ [GeV] & $M_3$ [GeV]\\
\hline
\hline
$6.95\times10^{-19}$ & $1.57\times10^{-17}$ & $2.23\times10^{-12}$ &
$3\times10^5$ & $8\times10^9$ & $2\times10^{16}$ \\
\hline
\end{tabular}
\end{center}
\caption{Typical magnitudes of branching ratios for lepton flavor violating
rare decay process.}
\label{tab11}
\end{table}
In these analyses, $\LamGUT=2\times10^{16}$ GeV is taken. 
These results are compared with the current experimental upper bounds 
at $90\%$ confidence level~\cite{Brooks:1999pu,Hayasaka:2007vc}: 
\begin{eqnarray}
  \mbox{Br}(\mu\rightarrow e\gamma)\leq1.2\times10^{-11},~~
  \mbox{Br}(\tau\rightarrow e\gamma)\leq1.2\times10^{-7},~~
  \mbox{Br}(\tau\rightarrow\mu\gamma)\leq4.5\times10^{-8}.
\end{eqnarray}
The magnitudes of the branching ratios for the lepton flavor violating 
process in the model with the applicable heaviest right-handed 
Majorana neutrino mass are far below the experimental limit. Once one 
fix the value of $\tan\beta$, the current experimental limit gives 
lower bound on $m_{\text{SUSY}}$, which is 
e.g. $m_{\text{SUSY}}\geq41.6$ GeV from 
$\mbox{Br}(\tau\rightarrow\mu\gamma)\leq4.5\times10^{-8}$ with 
$\tan\beta=38$. In the present case, the bound can be easily satisfied. 

\subsection{Leptogenesis}

At the end of this section, we examine whether the thermal 
leptogenesis~\cite{Fukugita:1986hr} works in our model. The CP 
asymmetry parameter in the decay process of right-handed neutrino, 
$R_i$, is given by 
\begin{eqnarray}
  \epsilon_i=\frac{\sum_j\Gamma(R_i\rightarrow L_jH)
    -\sum_j\Gamma(R_i\rightarrow L_j^cH^\dagger)}
  {\sum_j\Gamma(R_i\rightarrow L_jH)
    +\sum_j\Gamma(R_i\rightarrow L_j^cH^\dagger)},
\end{eqnarray}
where $L_i$ and $H$ denote the left-handed lepton and Higgs fields. An 
approximation for $\epsilon_i$ at low temperature is estimated 
by~\cite{LPG}, 
$\epsilon_1=\frac{1}{8\pi}\sum_{i\neq1}\mbox{Im}[(A_{i1})^2]F(r_i)/|A_{11}|$, 
where $r_i\equiv|M_i/M_1|^2$, $A\equiv(DM_{\nu D}M_{\nu 
  D}^{\dagger}D^\dagger)/v_u^2$, and $D$ being a diagonal phase matrix 
to make the eigenvalues $M_i$ real and positive. The function $F$ 
describes contributions from the one-loop vertex and self-energy 
corrections, 
\begin{eqnarray}
  F(x)=\sqrt{x}\left[\frac{2}{1-x}-\ln\left(1+\frac{1}{x}\right)\right].
\end{eqnarray}
We here define the resultant CP asymmetry, $\etaCP$, as the ratio of
the lepton asymmetry to the photon number density $n_\gamma$,
$\etaCP=135\zeta(3)\kappa s\epsilon_1/(4\pi^4g_\ast n_\gamma)$. In 
the equation, $\kappa$, $s$, and $g_\ast$ are the efficiency factor, 
entropy density, and the effective number of degrees of freedom in 
thermal equilibrium. They are given by~\cite{LPG2}, $s=7.04n_\gamma$, 
$g_\ast=228.75$, and
\begin{eqnarray}
  \kappa^{-1} &\simeq& \frac{3.3\times10^{-3}\mbox{ eV}}{\meff}
  +\left(\frac{\meff}{5.5\times10^{-4}\mbox{ eV}}
  \right)^{1.16}.
\end{eqnarray}
The $\meff$ is the effective light neutrino mass defined as 
$\meff\equiv|(M_{\nu D}^{\dagger}M_{\nu D})_{11}/M_1|$. The BAU, 
$\eta_B$, is transferred via spharelon interactions as 
$\eta_B=-8\etaCP/23$. Finally, the baryon asymmetry in our model is 
predicted as $\eta_B\sim10^{-23}\sin\theta_B$, where 
$\theta_B\equiv\theta_3-\theta_1$ and $\theta_i=\arg(M_i)$. These 
results are compared with the current observational data at $68\%$ 
confidence level from the WMAP 7-year resulting in the standard 
$\Lambda$CDM model~\cite{Komatsu:2010fb}. We can see that the baryon 
asymmetry generated through the leptogenesis is too small to explain 
the BAU. This is because the hierarchy in the Dirac neutrino mass matrix 
in the $SO(10)$ model is determined by the up-type quark mass matrix. 
Therefore, there is no freedom to adjust the Dirac neutrino mass matrix 
such that the BAU can be generated in the present model. It is 
expected that enough BAU might be realized by extending our cascade 
model to the inverse seesaw case, see~\cite{inverseseesaw}. In the 
case, the structure of effective light neutrino mass matrix is 
slightly changed but the realistic PMNS mixing angels would be 
obtained. 
\section{Discussion}

At the end of the paper, we give a comment on phenomenological aspects 
of proton decay. In general for SUSY GUTs, there are three sources 
that mediate the proton decays. The first one comes from the 
dimension-6 operators, arising from the exchanging of the heavy gauge 
bosons. Note that this type of operators exists in both non-SUSY and 
SUSY GUTs. These operators are significantly suppressed by 
$1/\Lambda_{\text{GUT}}^2$, therefore, we have no problem with the proton decay
 from these operators if $\Lambda_{\text{GUT}}$ is large enough, i.e. 
$\Lambda_{\text{GUT}}\geq10^{16}$~GeV. The second source is from the 
dimension-5 operators, arising from the exchanging of color triplet Higgsino 
fields. In this case, the proton decay contribution is suppressed by $1/M_{H}$,
 where $M_{H}$ is the mass of the color triplet Higgsinos. To suppress the 
proton decay contributions, the mass $M_{H}$ has to be very heavy, which can 
achieved by some doublet-triplet splitting 
mechanism~\cite{Dimopoulos:1981xm,Chacko:1998zn,Lee:1994je}. The third 
contribution arises from the dimension-4 operators, which are not suppressed by
 the GUT scale, however, these operators are eliminated by the $R-$parity. In 
the class of $SO(10)$ models, which do not contain the spinor $16$ or 
$\bar{16}$ as the Higgs fields, the $R-$parity is conserved. Otherwise, in 
order to avoid the proton decay contributions from these operators, the 
$R-$parity has to be introduced by hands. Since we consider an $SO(10)$ model 
without using the spinor Higgs representations, we have no problem with these 
dimension-4 operators~\cite{Mohapatra:1986su,Martin:1992mq}. 

\section{Summary}

We have done texture analyses of cascade model in supersymmetric 
$SO(10)$ model. The neutrino Dirac mass matrix of a cascade form can 
realize the tri-bimaximal mixing at the leading order while the 
down-quark one of a H.C. form can lead to realistic structure of CKM 
mixing. This fact gives us a strong motivation to study cascade 
hierarchical textures in a grand unified theory. 

We analytically clarified possible structures of neutrino Dirac, 
charged lepton, quark, and right-handed neutrino mass matrices by 
estimating collection from them to the tri-bimaximal mixing. The 
numerical analyses based on two typical models have been also 
presented. The minimal cascade texture in the context of SUSY $SO(10)$ 
GUT can lead to clear predictions for the PMNS mixing angles, which 
are $0.29\lesssim\sin^2\theta_{12}$ and 
$0.002\lesssim\sin^2\theta_{13}\lesssim0.007$, and relatively sharp 
correlations between the reactor and atmospheric angles. This result 
is a hot topic for the upcoming experiments of the reactor neutrino 
mixing angle. It might be checked by such future experiments. We have 
also shown that our typical cascade models can pass the constraints 
from the lepton flavor violation searches. For generating the BAU, we 
cannot generate enough asymmetry through the thermal leptogenesis 
mechanism in our model. Therefore, we need other mechanisms to 
generate the BAU. 

\subsection*{Acknowledgement}

The work of A.A. and R.T. is supported by the DFG-SFB TR 27.

\appendix

\section{Constraints on structure of non-diagonal {\boldmath $M_R$ case}}

The constraints on the structure of non-diagonal $M_R$ are presented 
in this Appendix. We defined the diagonalized mass matrix of the 
right-handed neutrino in~(\ref{DR}) and an unitary matrix which 
diagonalizes the $M_R$. The neutrino mass after the seesaw mechanism 
and operating the $V_{\text{TB}}$ is given in~\eqref{neu-mass2}. All 
matrix elements are given by 
\begin{eqnarray}
  \mathcal{M}_{11}
  &\simeq& \frac{
    v_u^2}{6M}
  [1+4\lambda^{16-x_2}+4\lambda^8\theta_{R,23}
  +\lambda^{-x_2}\theta_{R,23}^2 \nonumber \\
  &      & \phantom{\frac{
      v_u^2}{6M}[}
  +\lambda^{-x_1}(4\lambda^{16}\theta_{R,12}
  -4\lambda^8\theta_{R,12}\theta_{R,13}+\theta_{R,13}^2)], \\
  \mathcal{M}_{22}
  &\simeq& \frac{
    v_u^2}{M}
  [3\lambda^{16-x_1}+\frac{1}{3}+\frac{\lambda^{16-x_2}}{3}
  +\lambda^{-x_2}(-\frac{2\lambda^8}{3}\theta_{R,23}+\theta_{R,23}^2)
  +\lambda^{-x_1}(-2\lambda^8\theta_{R,13}+\theta_{R,13}^2)],\nonumber
  \\ && \\
  \mathcal{M}_{33}
  &\simeq& \frac{
    v_u^2}{M}
  [2\lambda^{8-x_2}+\frac{1}{2}
  +\lambda^{-x_2}(2\lambda^4\theta_{R,23}+\frac{\theta_{R,23}^2}{2})
  -2\lambda^{4-x_1}\theta_{R,12}\theta_{R,13}
  \nonumber \\
  &      & \phantom{\frac{
      v_u^2}{M}[}
  +\lambda^{-x_1}(2\lambda^8\theta_{R,12}^2
  +\frac{1}{2}\theta_{R,13}^2)], \\
  \mathcal{M}_{12}
  &\simeq& -\frac{
    v_u^2}{3\sqrt{2}M}
  [1+\lambda^{8-x_2}\theta_{R,23}
  +3\lambda^{-x_1}(-2\lambda^{16}\theta_{R,12}
  +\lambda^8\theta_{R,13})], \\
  \mathcal{M}_{23}
  &\simeq& \frac{
    v_u^2}{\sqrt{6}M}
  [1-2\lambda^{12-x_2}+2\lambda^{4-x_2}\theta_{R,23} \nonumber \\
  &      & \phantom{\frac{
      v_u^2}{\sqrt{6}M}[}
  +3\lambda^{-x_1}(2\lambda^{12}\theta_{R,12}-\lambda^8\theta_{R,13}
  -2\lambda^4\theta_{R,13}\theta_{R,12}
  +\theta_{R,13}^2)], \\
  \mathcal{M}_{13}
  &\simeq& -\frac{
    v_u^2}{2\sqrt{3}M}
  [1+\lambda^{-x_2}(4\lambda^{12}+2\lambda^4\theta_{R,23}
  +\theta_{R,23}^2) \nonumber \\
  &      & \phantom{-\frac{
      v_u^2}{2\sqrt{3}M}[}
  -\lambda^{-x_1}(2\lambda^4\theta_{R,12}\theta_{R,13}
  -4\lambda^{12}\theta_{R,12}^2-\theta_{R,13}^2)].
\end{eqnarray}
We require that the magnitudes of leading order of each term in this 
mass matrix are the same one as in the case of diagonal $M_R$ case 
because the tri-bimaximal mixing can be already realized at the 
leading order by neutrino Dirac mass matrix. It leads to constraints 
on the mixing angles as follows: 
\begin{eqnarray}
  \theta_{R,13} &<& \frac{3}{2}\lambda^8,~\frac{1}{3}\lambda^{-8+x_1},~
  2\lambda^{4+(x_1-x_2)/2},~\frac{1}{\sqrt{3}}\lambda^{x_1/2},
  \\
  \theta_{R,23} &<&
  \lambda^4,~\frac{1}{2}\lambda^{-4+x_2},~\lambda^{x_2/2} \\
  \theta_{R,12} &<&
  \frac{1}{6}\lambda^{-12+x_1},~\lambda^{(x_1-x_2)/2},~\frac{1}{2}\lambda^{-6+x_1/2},
  \\
  \theta_{R,12}\theta_{R,13} &<&
  \lambda^{4+x_1-x_2},~\frac{1}{6}\lambda^{-4+x_1}.
\end{eqnarray}
After fixing the values of $(d_1,d_2,x_1,x_2)$ so that they must 
satisfy the four conditions, one can obtain the structure leading to 
maximal collections to the PMNS mixing angles and neutrino mass 
spectra as shown in Tabs.~\ref{tab9} and~\ref{tab10}. 



\begin{thebibliography}{99}
\bibitem{Schwetz:2008er}
  T.~Schwetz, M.~Tortola and J.~W.~F.~Valle,
  New J.\ Phys.\  {\bf 10} (2008) 113011.
  
\bibitem{TBM}
  P.F.~Harrison, D.H.~Perkins and W.G.~Scott,
  Phys.~Lett. {\bf B 530} (2002) 167;
  P.F.~Harrison and W.G.~Scott,
  Phys.~Lett. {\bf B 535} (2002) 163.
  
\bibitem{Altarelli:2005yp}
  G.~Altarelli and F.~Feruglio,
  Nucl.\ Phys.\  B {\bf 720} (2005) 64;
  S.~F.~King,
  JHEP {\bf 0508} (2005) 105;
  I.~de Medeiros Varzielas and G.~G.~Ross,
  Nucl.\ Phys.\  B {\bf 733} (2006) 31;
  E.~Ma,
  Phys.\ Lett.\  B {\bf 632} (2006) 352;
  A.~Zee,
  Phys.\ Lett.\  B {\bf 630} (2005) 58;
  W.~Grimus and L.~Lavoura,
  JHEP {\bf 0601} (2006) 018;
  E.~Ma,
  Phys.\ Rev.\  D {\bf 73} (2006) 057304;
  G.~Altarelli and F.~Feruglio,
  Nucl.\ Phys.\  B {\bf 741} (2006) 215;
  J.~E.~Kim and J.~C.~Park,
  JHEP {\bf 0605} (2006) 017;
  I.~de Medeiros Varzielas, S.~F.~King and G.~G.~Ross,
  Phys.\ Lett.\  B {\bf 644} (2007) 153;
  R.~N.~Mohapatra, S.~Nasri and H.~B.~Yu,
  Phys.\ Lett.\  B {\bf 639} (2006) 318;
  I.~de Medeiros Varzielas, S.~F.~King and G.~G.~Ross,
  Phys.\ Lett.\  B {\bf 648} (2007) 201;
  E.~Ma,
  Mod.\ Phys.\ Lett.\  A {\bf 21} (2006) 2931;
  G.~Altarelli, F.~Feruglio and Y.~Lin,
  Nucl.\ Phys.\  B {\bf 775} (2007) 31;
  H.~Zhang,
  Phys.\ Lett.\  B {\bf 655} (2007) 132;
  P.~D.~Carr and P.~H.~Frampton,
  [arXiv:hep-ph/0701034];
  F.~Feruglio, C.~Hagedorn, Y.~Lin and L.~Merlo,
  Nucl.\ Phys.\  B {\bf 775} (2007) 120
  [Erratum-ibid.\  {\bf 836} (2010) 127];
  M.~C.~Chen and K.~T.~Mahanthappa,
  Phys.\ Lett.\  B {\bf 652} (2007) 34;
  C.~Luhn, S.~Nasri and P.~Ramond,
  Phys.\ Lett.\  B {\bf 652} (2007) 27;
  Y.~Koide,
  arXiv:0707.0899 [hep-ph];
  E.~Ma,
  Phys.\ Lett.\  B {\bf 660} (2008) 505;
  F.~Bazzocchi, S.~Morisi and M.~Picariello,
  Phys.\ Lett.\  B {\bf 659} (2008) 628;
  F.~Plentinger, G.~Seidl and W.~Winter,
  JHEP {\bf 0804} (2008) 077;
  F.~Plentinger and G.~Seidl,
  Phys.\ Rev.\  D {\bf 78} (2008) 045004;
  S.~Antusch, S.~F.~King and M.~Malinsky,
  JHEP {\bf 0805} (2008) 066;
  Y.~Lin,
  Nucl.\ Phys.\  B {\bf 813} (2009) 91;
  F.~Feruglio, C.~Hagedorn, Y.~Lin and L.~Merlo,
  Nucl.\ Phys.\  B {\bf 809} (2009) 218;
  T.~Araki and R.~Takahashi,
  Eur.\ Phys.\ J.\  C {\bf 63} (2009) 521;
  W.~Grimus and L.~Lavoura,
  JHEP {\bf 0904} (2009) 013;
  H.~Ishimori, Y.~Shimizu and M.~Tanimoto,
  Prog.\ Theor.\ Phys.\  {\bf 121} (2009) 769;
  S.~Morisi,
  Phys.\ Rev.\  D {\bf 79} (2009) 033008;
  F.~Bazzocchi, L.~Merlo and S.~Morisi,
  Nucl.\ Phys.\  B {\bf 816} (2009) 204;
  K.~Kojima and H.~Sawanaka,
  Phys.\ Lett.\  B {\bf 678} (2009) 373;
  F.~Bazzocchi, L.~Merlo and S.~Morisi,
  Phys.\ Rev.\  D {\bf 80} (2009) 053003;
  A.~Hayakawa, H.~Ishimori, Y.~Shimizu and M.~Tanimoto,
  Phys.\ Lett.\  B {\bf 680} (2009) 334;
  G.~Altarelli and D.~Meloni,
  J.\ Phys.\ G {\bf 36} (2009) 085005;
  M.~Hirsch, S.~Morisi and J.~W.~F.~Valle,
  Phys.\ Lett.\  B {\bf 679} (2009) 454;
  Y.~Lin,
  Nucl.\ Phys.\  B {\bf 824} (2010) 95;
  A.~Adulpravitchai, M.~Lindner and A.~Merle,
  Phys.\ Rev.\  D {\bf 80} (2009) 055031;
  A.~Adulpravitchai, M.~Lindner, A.~Merle and R.~N.~Mohapatra,
  Phys.\ Lett.\  B {\bf 680} (2009) 476;
  D.~Aristizabal Sierra, F.~Bazzocchi, I.~de Medeiros Varzielas, L.~Merlo and S.~Morisi,
  Nucl.\ Phys.\  B {\bf 827} (2010) 34;
  T.~J.~Burrows and S.~F.~King,
  Nucl.\ Phys.\  B {\bf 835} (2010) 174;
  F.~Feruglio, C.~Hagedorn and L.~Merlo,
  JHEP {\bf 1003} (2010) 084;
  B.~Dutta, Y.~Mimura and R.~N.~Mohapatra,
  JHEP {\bf 1005} (2010) 034;
  Y.~Lin, L.~Merlo and A.~Paris,
  Nucl.\ Phys.\  B {\bf 835} (2010) 238;
  F.~Feruglio, C.~Hagedorn, Y.~Lin and L.~Merlo,
  arXiv:0911.3874 [hep-ph];
  A.~Adulpravitchai and M.~A.~Schmidt,
  arXiv:1001.3172 [hep-ph];
  C.~Hagedorn, S.~F.~King and C.~Luhn,
  JHEP {\bf 1006} (2010) 048;
  H.~Ishimori, K.~Saga, Y.~Shimizu and M.~Tanimoto,
  Phys.\ Rev.\  D {\bf 81} (2010) 115009;
  G.~J.~Ding,
  arXiv:1006.4800 [hep-ph];
  T.~J.~Burrows and S.~F.~King,
  Nucl.\ Phys.\  B {\bf 842} (2011) 107;
  Y.~Shimizu and R.~Takahashi,
  arXiv:1009.5504 [hep-ph];
  T.~Araki, J.~Mei and Z.~z.~Xing,
  arXiv:1010.3065 [hep-ph].
  
\bibitem{Haba:2008dp}
  N.~Haba, R.~Takahashi, M.~Tanimoto and K.~Yoshioka,
  Phys.\ Rev.\  D {\bf 78} (2008) 113002.

\bibitem{su5}
  K.~Kojima, H.~Sawanaka and R.~Takahashi,
  arXiv:1011.5678 [hep-ph].
  
\bibitem{Georgi:1979df}
  H.~Georgi and C.~Jarlskog,
  Phys.\ Lett.\  B {\bf 86} (1979) 297.

\bibitem{Aulakh:1982sw}
  C.~S.~Aulakh, R.~N.~Mohapatra,
  Phys.\ Rev.\  {\bf D28 } (1983)  217.

\bibitem{Clark:1982ai}
  T.~E.~Clark, T.~-K.~Kuo, N.~Nakagawa,
  Phys.\ Lett.\  {\bf B115 } (1982)  26.

\bibitem{Aulakh:2003kg}
  C.~S.~Aulakh, B.~Bajc, A.~Melfo {\it et al.},
  Phys.\ Lett.\  {\bf B588 } (2004)  196-202.

\bibitem{Bajc:2004xe}
  B.~Bajc, A.~Melfo, G.~Senjanovic {\it et al.},
  Phys.\ Rev.\  {\bf D70 } (2004)  035007.


\bibitem{Chen:2003zv}
  M.~C.~Chen and K.~T.~Mahanthappa,
  Int.\ J.\ Mod.\ Phys.\  A {\bf 18} (2003) 5819.

\bibitem{Dimopoulos:1981xm}
  S.~Dimopoulos and F.~Wilczek, Print-81-0600 (SANTA BARBARA)

\bibitem{Chacko:1998zn}
  Z.~Chacko and R.~N.~Mohapatra,
  Phys.\ Rev.\ Lett.\  {\bf 82} (1999) 2836.

\bibitem{Lee:1994je}
  D.~G.~Lee and R.~N.~Mohapatra,
  Phys.\ Rev.\  D {\bf 51} (1995) 1353.

\bibitem{Fukuyama:2004ps}
  T.~Fukuyama, A.~Ilakovac, T.~Kikuchi, S.~Meljanac and N.~Okada,
  J.\ Math.\ Phys.\  {\bf 46} (2005) 033505.

\bibitem{Ross:2007az}
  G.~Ross and M.~Serna,
  Phys.\ Lett.\  B {\bf 664} (2008) 97.

\bibitem{Carena:1999py}
  R.~Hempfling,
  Phys.~Rev. {\bf D49} (1994) 6168;
  L.J.~Hall, R.~Rattazzi and U.~Sarid,
  Phys.~Rev. {\bf D50} (1994) 7048;
  M.~Carena, M.~Olechowski, S.~Pokorski and C.E.M.~Wagner,
  Nucl.~Phys. {\bf B426} (1994) 269;
  T.~Blazek, S.~Raby and S.~Pokorski,
  Phys.~Rev. {\bf D52} (1995) 4151;
  D.M.~Pierce, J.A.~Bagger, K.T.~Matchev and R.j.~Zhang,
  Nucl.~Phys. {\bf B491} (1997) 3;
  M.~S.~Carena, D.~Garcia, U.~Nierste and C.~E.~M.~Wagner,
  Nucl.\ Phys.\  B {\bf 577} (2000) 88;
  M.~S.~Carena and H.~E.~Haber,
  Prog.\ Part.\ Nucl.\ Phys.\  {\bf 50} (2003) 63;
  K.~Tobe and J.~D.~Wells,
  Nucl.\ Phys.\  B {\bf 663} (2003) 123;
  K.~Inoue, K.~Kojima and K.~Yoshioka,
  JHEP {\bf 0607} (2006) 032 ;
  Phys.\ Lett.  {\bf B644} (2007) 172.

\bibitem{Ardellier:2006mn}
  F.~Ardellier {\it et al.}  [Double Chooz Collaboration],
  arXiv:hep-ex/0606025.

\bibitem{:2010vy}
  J.~K.~Ahn  [RENO Collaboration],
  arXiv:1003.1391 [hep-ex].

\bibitem{Guo:2007ug}
  X.~Guo {\it et al.}  [Daya-Bay Collaboration],
  arXiv:hep-ex/0701029.

\bibitem{Itow:2001ee}
  Y.~Itow {\it et al.}  [The T2K Collaboration],
  arXiv:hep-ex/0106019.

\bibitem{Ayres:2004js}
  D.~S.~Ayres {\it et al.}  [NOvA Collaboration],
  arXiv:hep-ex/0503053.

\bibitem{Mezzetto:2010zi}
  M.~Mezzetto and T.~Schwetz,
  arXiv:1003.5800 [hep-ph].

\bibitem{Dutta:2009ij}
  B.~Dutta, Y.~Mimura, R.~N.~Mohapatra,
  Phys.\ Rev.\  {\bf D80 } (2009)  095021.

\bibitem{Borzumati:1986qx}
  F.~Borzumati and A.~Masiero,
  Phys.\ Rev.\ Lett.\  {\bf 57} (1986) 961;
  J.~Hisano, T.~Moroi, K.~Tobe and M.~Yamaguchi,
  Phys.\ Rev.\  D {\bf 53} (1996) 2442;
  J.~R.~Ellis, M.~E.~Gomez, G.~K.~Leontaris, S.~Lola and D.~V.~Nanopoulos,
  Eur.\ Phys.\ J.\  C {\bf 14} (2000) 319;

\bibitem{Brooks:1999pu}
  M.~L.~Brooks {\it et al.}  [MEGA Collaboration],
  Phys.\ Rev.\ Lett.\  {\bf 83} (1999) 1521.

\bibitem{Hayasaka:2007vc}
  K.~Hayasaka {\it et al.}  [Belle Collaboration],
  Phys.\ Lett.\  B {\bf 666} (2008) 16.

\bibitem{Fukugita:1986hr}
  M.~Fukugita and T.~Yanagida,
  Phys.\ Lett.\  B {\bf 174} (1986) 45.

\bibitem{LPG}
  L.~Covi, E.~Roulet and F.~Vissani,
  Phys.~Lett. {\bf B384} (1996) 169;
  W.~Buchmuller and M.~Plumacher,
  Phys.~Lett. {\bf B389} (1996) 73;
  A.~Pilaftsis,
  Phys.~Rev. {\bf D56} (1997) 5431.

\bibitem{LPG2}
  G.F.~Giudice, A.~Notari, M.~Raidal, A.~Riotto and A.~Strumia,
  Nucl.~Phys. {\bf B685} (2004) 89.
  
\bibitem{Komatsu:2010fb}
  E.~Komatsu {\it et al.},
  arXiv:1001.4538 [astro-ph.CO].

\bibitem{inverseseesaw}
  S.~Blanchet, P.~S.~B.~Dev and R.~N.~Mohapatra,
  arXiv:1010.1471 [hep-ph].


\bibitem{Mohapatra:1986su}
  R.~N.~Mohapatra,
  Phys.\ Rev.\  D {\bf 34} (1986) 3457.
  
\bibitem{Martin:1992mq}
  S.~P.~Martin,
  Phys.\ Rev.\  D {\bf 46} (1992) 2769.
\end{thebibliography}
\end{document}